\documentclass[twocolumn]{article}

\usepackage[utf8x]{inputenc}
\usepackage[english]{babel}
\usepackage[T1]{fontenc}

\usepackage[table]{xcolor}
\usepackage{graphicx}
\usepackage{subfigure}
\usepackage{float}
\usepackage{amsmath}
\usepackage{amssymb}
\usepackage{enumerate}
\usepackage{fancybox}
\usepackage{epstopdf}
\usepackage{hyperref}
\usepackage{cases}

\usepackage{array}
\usepackage{multirow}
\usepackage{booktabs}
\usepackage{setspace}
\usepackage{appendix}

\newcounter{tempEquationCounter}
\newcounter{thisEquationNumber}

\hypersetup{
backref=true,            
pagebackref=true,    
hyperindex=true,      
colorlinks=true,        
breaklinks=true,       
urlcolor= orange,        
linkcolor= blue,        
citecolor=red,
bookmarks=true,       
bookmarksopen=false,  
}

\newcommand{\rem}[1]{}

\usepackage{abstract}
	
\title{Analytical determination of the attack transient in a clarinet with time-varying blowing pressure}

\author{A. Almeida$^{a,b,}$\thanks{Corresponding author: \href{mailto:andre.almeida@univ-lemans.fr}{andre.almeida@univ-lemans.fr}}, B. Bergeot$^{a,c}$, C. Vergez$^{c}$ and B. Gazengel$^{a}$}

\date{{\small \textit{$^{a}$LUNAM Universit\'{e}, Universit\'{e} du Maine, UMR CNRS 6613, Laboratoire d’Acoustique, Avenue Olivier Messiaen, 72085 Le Mans Cedex 9, France}}\\
{\small \textit{$^{b}$ School of Physics, The University of New South Wales, Sydney UNSW 2052, Australia}}\\
{\small \textit{$^{c}$LMA, CNRS UPR7051, Aix-Marseille Univ., Centrale Marseille, F-13402 Marseille Cedex 20, France}}}

\begin{document}

\maketitle

\begin{abstract}
  This article uses a basic model of a reed instrument,
  known as the lossless Raman model, to determine analytically the
  envelope of the sound produced by the clarinet
  when the mouth pressure is increased gradually to start a
  note from silence. Using results from
  dynamic bifurcation theory, a prediction of the amplitude of the
  sound as a function of time is given
  based on a few parameters quantifying the time evolution of mouth
  pressure. As in previous uses of this model, the predictions are
  expected to be qualitatively consistent with
  simulations using the Raman model, and observations of real instruments.
  Model simulations for slowly variable parameters
  require very high precisions of computation. Similarly, any real
  system, even if close to the model would be affected by noise. In order to describe
  the influence of noise, a modified model is developed that includes a
  stochastic variation of the parameters.
Both ideal and stochastic models are shown to attain a minimal
amplitude at the static oscillation threshold. Beyond this point, the
amplitude of the oscillations increases exponentially, although some
time is required before the oscillations can be observed at the
``dynamic oscillation threshold''.
The effect of a sudden interruption of the growth of the mouth
pressure is also studied, showing that it usually triggers a faster growth of
the oscillations.
\end{abstract}

\section{Introduction}
\label{sec:introductionRe}


One of the many skills involved in learning how to play the clarinet
is to control the attack of a new note.  Tonguing is an important
aspect of a clear and precise attack, but the evolution of the mouth
pressure during the first instants of the note is also seen to affect
the attack considerably. Moreover, in some particular situations, tonguing may
not be involved in starting a new note. Hence there is both scientific
and practical interest in the question: what combinations of tonguing and evolution of
the blowing pressure produces sharp and precise attacks?

As a self-sustained musical instrument, the clarinet can be seen as a
dynamic system in which the oscillation is controlled by input
parameters from the musician. Two of the most important
\cite{McIntyre83:JASA,WilJASA1974,schum1981} are the blowing pressure and
the lip force upon the reed. Models predict the range of
parameter values that
allow for the production of a musical note
\cite{OllivActAc2005}. Other useful predictions are the dependence of
amplitude of oscillation on these two parameters, period doubling
bifurcation points \cite{KergoCFA2004,NonLin_Tail_2010}, or the
parameter regions where the reed touches the lay of the mouthpiece.
\cite{KergoActa2000,dalmont:3294}.


More complex models exist that can simulate the reed oscillation in time-domain \cite{Keefe1992}
or harmonic balance methods \cite{farner2006contribution}. They provide more
accurate predictions but their complexity makes it hard to grasp the
causality relation between parameters and consequences in oscillatory behaviour.

In previous studies in which mouth pressure was gradually increased at
constant rates, oscillations appeared at a much higher mouth pressure
threshold than that predicted assuming a constant mouth
pressure. High thresholds were observed in an artificially blown clarinet
\cite{Jasa2013BBergeot} and even higher in numerical simulations \cite{ThFabSil}.



Analytical reasoning ~\cite{BergeotNLD2012} based on dynamic bifurcation theory
\cite{Baesens1991,FruchaScaf2003} predicts a delay in the threshold of
oscillation for a linearly increasing mouth pressure, but the exact value of mouth pressure
at which it occurs is only valid for simulations performed with very
high precision. The threshold observed with normal
precision simulations can only be explained with a modified theory
\cite{BergeotNLD2012b} using stochastic perturbations \cite{Baesens1991}.


This article extends previous studies by the present authors by switching
the focus from the threshold of oscillation to a complete description
of the amplitude of oscillation. A simplified model of a note
attack is a constant increase in the mouth pressure (as used in
previous articles) which ceases increasing and then remains constant
at a defined value. The effect of ceasing the pressure increase
is studied analytically to develop a full recipe for estimating the
envelope of the attack. This recipe is then explored by
comparing to actual simulations of the Raman model.



In section \ref{sec:clari-theo}, the model of the clarinet used in
this work is briefly presented, as well as some of its known
properties. The remaining of this section provides a brief overview of
the key concepts that are needed for the present article (most of these
concepts are described with more details in two articles by the authors
\cite{BergeotNLD2012, BergeotNLD2012b}).
Section \ref{sec:envel-after-dynam} describes the calculation of the
envelope of the oscillations relative to the invariant curve, firstly
in an ideal case with infinite precision, then with limited precision or noise
(section~\ref{sec:traj-syst-affect}). To some extent, these methods were already
employed in previous articles \cite{BergeotNLD2012,BergeotNLD2012b} to
determine a dynamic threshold of oscillation. Here they are extended
to calculate the envelope before this threshold is
reached. Section \ref{sec:interr-ramps-contr} presents a method to
take into account a discontinuity in the time derivative of the mouth
pressure. In section \ref{sec:examples}, the models are applied to
particular examples and simulations, analysing the consequences in
terms of expected evolution of the sound. A list of the symbols used in this article is provided in Appendix~\ref{sec:notation-table}.

\section{Elements of clarinet theory}
\label{sec:clari-theo}


\subsection{The clarinet model}
\label{sec:elementary-modelRe}

For an elementary analysis, the clarinet can be described using a
version of the
lossless Raman model \cite{OllivActAc2004}, originally used for the bowed
string. The system is described by two state variables $p$ and $u$,
made non-dimensional by dividing them respectively by the minimum
pressure that closes the reed in steady-state, and the maximum flow allowed by the
reed valve. A non-linear function $u=F(p)$ relates the pressure difference
between the mouth and the mouthpiece ($\Delta p = \gamma - p$, where
$\gamma$ is the mouth pressure) to the volume of air that flows
past the reed ($u$). The derivation of this formula is given for instance by Chaigne and Kergomard~\cite{Cha08Belin}.

 \begin{subnumcases}{\label{nonlin_carac_2eq_ad}F(p)=}
 \zeta \left(\Delta p - 1 \right)\sqrt{-\Delta p} \hspace{0.5cm}  \text{if} \ \Delta p <0; \\
 \zeta \left(1-\Delta p \right)\sqrt{\Delta p}  \hspace{0.35cm}
 \text{if} \ \Delta p \in [0,1]; \\
  0   \hspace{3.1cm} \text{if} \ \Delta p> 1.\label{carNL_beatreed}
 \end{subnumcases}

\begin{figure}[t]
\centering
\subfigure[$F$ function]{\includegraphics[width=0.8\columnwidth,keepaspectratio=true]{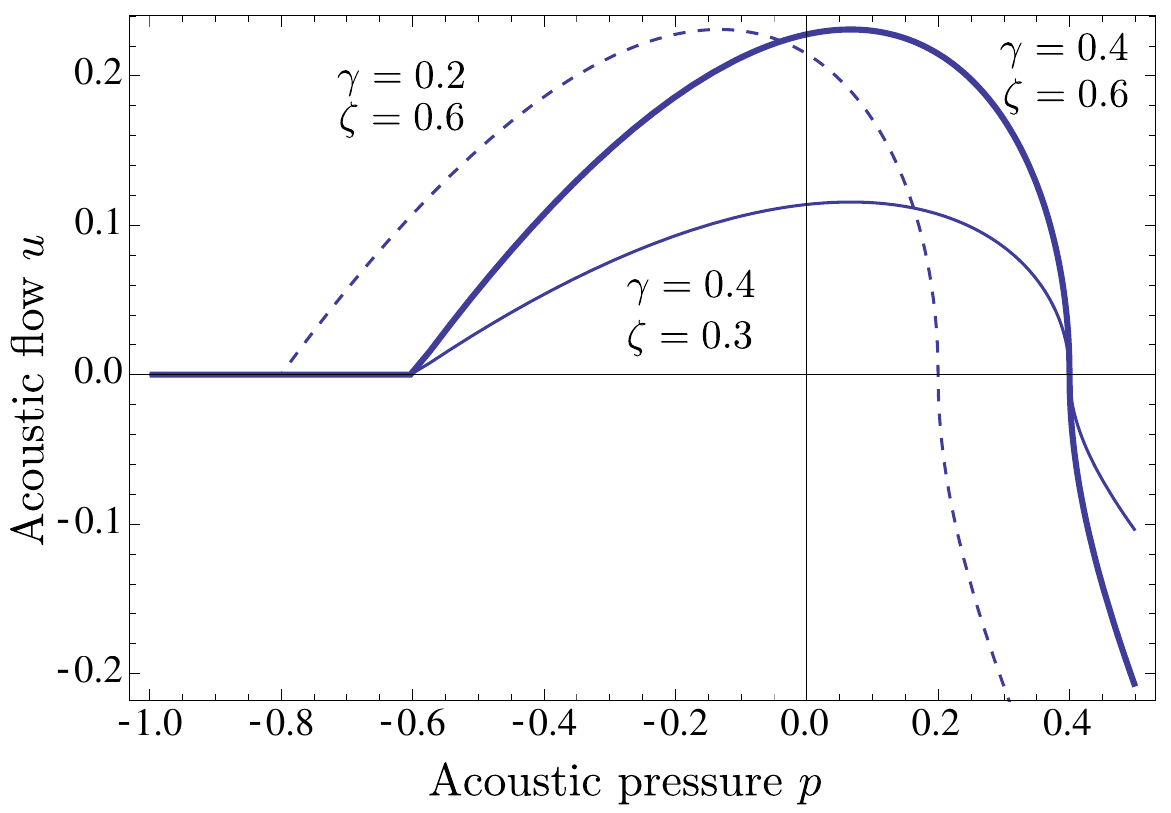}\label{rep_FRe}}
\subfigure[$G$ function]{\includegraphics[width=0.8\columnwidth,keepaspectratio=true]{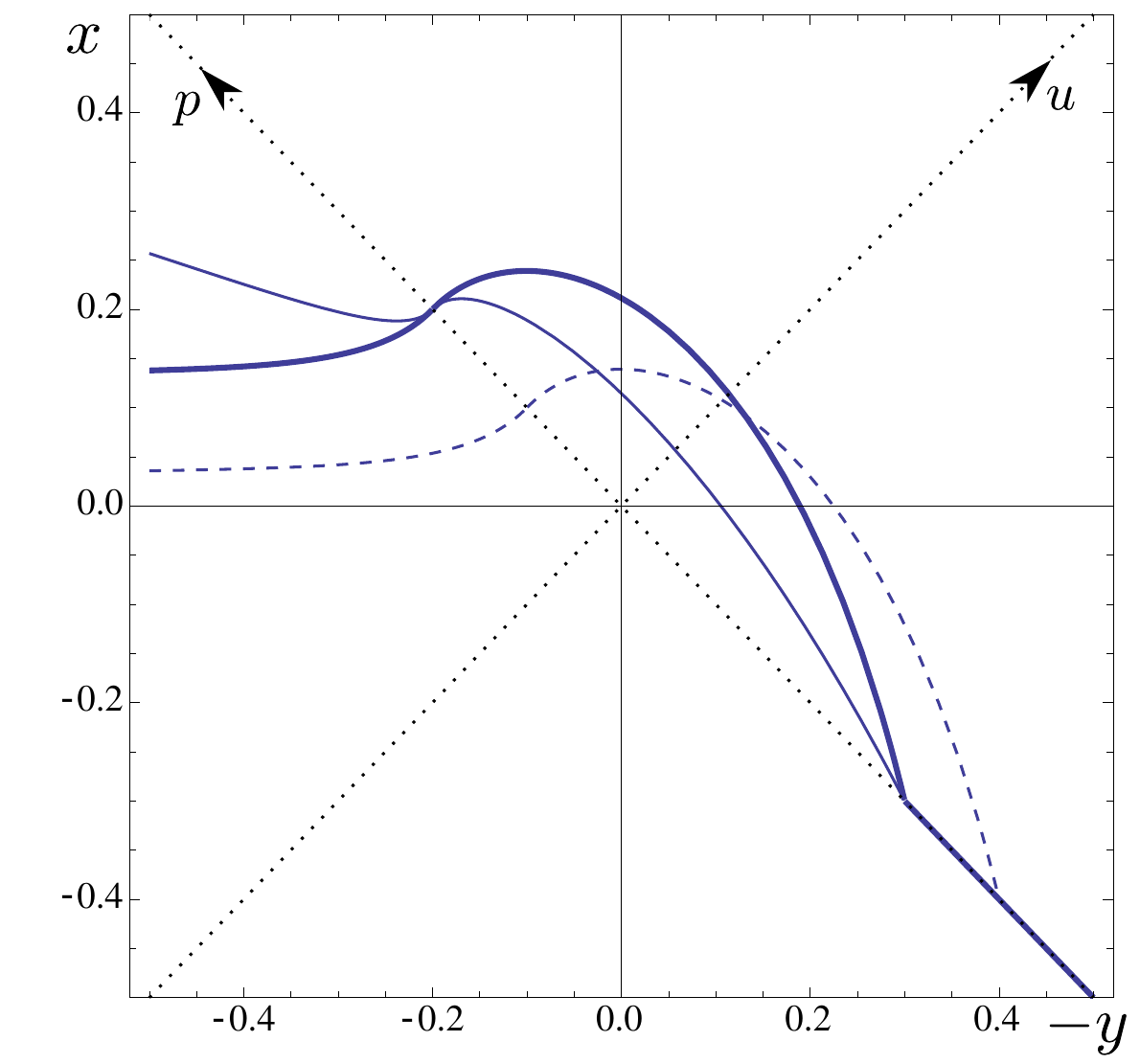}\label{rep_GRe}}
\caption{Non-linear characteristics in $u=F(p)$ representation (a) and
  $x=G(y)$ representation (b) for 3 different parameter values}
\end{figure}

The control parameters of the system are the mouth pressure $\gamma$
and the embouchre parameter $\zeta=\frac{\rho c}{S_{res}}
  S\sqrt{\frac{2P_M}{\rho}}\frac{1}{P_M}$. $\zeta$ is related the lip force
  via the opening area of the reed at rest $S$ and is proportional to the
  characteristic impedance at the resonator input $\frac{\rho
    c}{S_{res}}$. Three examples of the function $F$ (Fig.~\ref{rep_FRe}) show that smaller values of $\zeta$ bring the
characteristic function closer to that of a stopped pipe ($u=0$). Increasing $\gamma$ shifts the curve along the $p$-axis.


The reed-mouthpiece system drives the resonator. It is linked to it by
the acoustic variables $p$ and $u$ found in Eq.
\eqref{nonlin_carac_2eq_ad}. For a time-domain description it is usually
simpler to describe the resonator using two non-dimensional traveling wave variables
$x$ and $y$, respectively the outgoing and incoming pressure waves:
\begin{eqnarray}
p(t)=x(t)+y(t),\nonumber \\
u(t)=x(t)-y(t).
\label{eq:3}
\end{eqnarray}
The incoming wave $y(t)$ at the bore input is the opposite of the
delayed outgoing wave $-x(t-\tau)$, since no losses in the propagation
or reflection are considered\footnote{$x$ and $y$ are usually written
  respectively as $p^+$ and $p^-$. The latter form is used in
  this article for conciseness.}. In practice, only one value of
$x(t)$ is calculated in each round-trip of the wave, with a duration
of $\tau=2l/c$, where $l$ is the resonator length and $c$ the speed of
sound. All the variables can thus be discretized, $x_n$ meaning the
value of a variable $x$ at time $n\tau$.

The behaviour of the whole instrument then can be described in a single iterative equation:

\begin{equation}
x_n=G\left(x_{n-1},\gamma\right).
\label{DiffEq_G_InflGamm}
\end{equation}

Function $G$ can be obtained by replacing $p(t)$ and $u(t)$ in function
$F$ with Eq. \eqref{eq:3}. An explicit formulation for $G$ is given by Taillard \emph{et
  al.} \cite{NonLin_Tail_2010}, for $\zeta<1$.
Fig. \ref{rep_GRe} shows that the change from coordinates $(p,u)$ to
$(x,y)$ can be performed graphically as a mirror about the axis $p=0$
and a $45^\circ$ rotation about the origin. Like $F$, $G$ also depends
on the control parameters $\gamma$ and $\zeta$. To keep the notation simple, the
parameters will be omitted when constant. $\gamma$ will be included as
an argument to the function when it varies with time.

In most works on the clarinet, functions $F$ and $G$ are studied in a
static-parameter regime, referring to a case where the instrument is blown
at a constant pressure with a constant force applied on the lip. This
article focuses on a case where the mouth pressure $\gamma$ varies
over time, a situation is referred to hereafter as dynamic-parameter
regime, or simply dynamic regime. The graphics of Fig. \ref{rep_GRe} thus change over time.

\subsection{Invariant manifolds and non-oscillating solutions}
\label{sec:stat-regime-with}


In a static-parameter regime, there is a
value of $\gamma=\gamma_{st}$ establishing the transition between non-oscillating
and oscillating solutions. This is called the \emph{static oscillation
  threshold}.  Above this value, the clarinet system can oscillate,
and will indeed oscillate for most initial values $x_0$. However, for
particular sets of initial conditions (in the scope of this paper sets
of $\gamma_0$ and $x_0$), the solution is non-oscillating. \rem{CV:
  Quand on initialise sr le point fixe c'est plus que non-ocillating,
  c'est même non-evolving AA: je parle aussi du cas de $\gamma$
  variable, donc la solution ne va pas rester toujours à la même
  valeur... CV: Et inversement il ne suffit pas que la solution soit
  non-oscillante pour que la condition initialle apartienne à la
  variété invariante. AA je pense qu'on ne dit pas ça...} These sets
correspond to the ``invariant manifolds''. If $\gamma$ does not vary
with time, the invariant manifold is called a fixed point, as the
variable $x$ will remain constant ($x=x_0$). The fixed point $x^*$ can
be found by solving:

\begin{equation}
x^{*}=G\left(x^{*}\right).
\label{fixed_point_G_def}
\end{equation}

$x^*$ is a function of $\gamma$, $x^*=x^*(\gamma)$.

When $\gamma$ varies with time, the invariant manifold cannot
correspond to a single fixed point, but is also
time-dependent, corresponding to an ``invariant curve''. Perhaps
surprisingly, it is not the set of values $x^*(\gamma_n)$.
The invariant curve is defined as the set of values ($x$,
$\gamma$) such that during the planned time-variation of $\gamma$,
this set of values will always be followed, independently of the
particular value the system is initiated in. The following equation is
a defining condition for this curve:
\begin{equation}
\phi_\epsilon(\gamma)=G\left(\phi_\epsilon(\gamma-\epsilon),\gamma\right).
\label{eqdiff_1ArtRec}
\end{equation}

A method for calculating the invariant curve for the clarinet
system is given in a previous article
\cite{BergeotNLD2012}. In appendix \ref{sec:perturb} simpler
expressions for the invariant curve are given by using the
characteristic curve expressed as $u=F(p)$ instead of function $G$. The
invariant curve depends on how the parameter $\gamma$ varies in
time, i.~e., it is different for different rates of variation of
$\gamma$ (different $\epsilon$ values).

\subsection{Local stability of non-oscillating solutions}
\label{sec:stab-non-osc}

In both static and dynamic cases, the non-oscillating solutions can be
either stable or unstable, depending on the
behaviour of the system initialized close to the invariant manifold.

If initialized with a value $x_0$ close to a stable invariant
manifold, the state variable $x$ will approach it
exponentially. Conversely, the state variable is repelled
exponentially by an unstable manifold while in its vicinity. The
distance to a fixed point (in a static-parameter case and while $x_n$
is sufficiently close to the fixed point) is an exponential function
of time (expressed as iteration number $n$) \cite{Cha08Belin}:
\begin{equation}
x_n-x^*\approx (x_0-x^*) \left[G'\left(x^{*}\right)\right]^n.
\label{sol_stat}
\end{equation}
where $G'\left(x^{*}\right)$ is the derivative of the iterative function at
the fixed point. When this value exceeds $1$,  the fixed point is
unstable and the oscillation grows. Due to the non-linear nature of
the system, the oscillation cannot
grow forever, of course, and it stabilises in a periodic solution.


\begin{figure}[t]
\centering
\includegraphics[width=0.9\columnwidth,keepaspectratio=true]{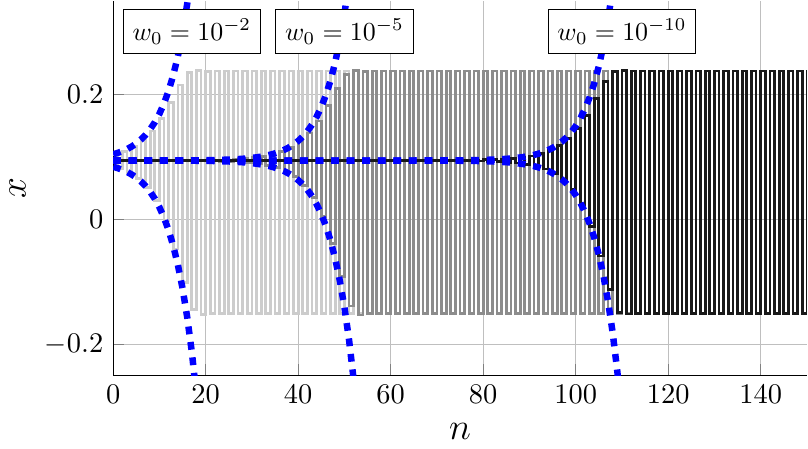}
\caption{Time evolution of the outgoing pressure $x$, solution of
  Eq.~\eqref{DiffEq_G_InflGamm} for different values of its
  initial value $x_0=x^*+w_0$. From left to right: $w_0=0.01$ \mbox{(\textcolor{gray}{\textbf{\color{black!20}-----}})},
$w_0=10^{-5}$
\mbox{(\textcolor{gray}{\textbf{\color{black!45}-----}})} and
$w_0=10^{-10}$
\mbox{(\textcolor{gray}{\textbf{\color{black!90}-----}})}. \mbox{(\textcolor{black}{\color{blue}\textbf{-
      - -}})} Exponential envelope deduced from the
function~\eqref{sol_stat}. The following parameters are used: $\gamma=0.42$ (constant) and $\zeta=0.5$.}
\label{figEDA:ExTrasCI1ArtRec1}
\end{figure}


In a static-parameter context, the oscillation would
eventually stabilise in an oscillatory regime between values given by
the 2-branch part of the static bifurcation diagram (an extensive discussion
is given by Taillard \emph{et al.} \cite{NonLin_Tail_2010}).

For time-varying parameters, the evolution of the system can be interpreted as a dynamic bifurcation diagram. In this case, it is observed that the system still follows
closely the invariant curve
$\phi_{\epsilon}$ even after it becomes unstable (see for instance
Fig.~\ref{zoom1}). Eventually an oscillation appears at a value of
mouth pressure much higher than the static oscillation threshold, so
that we speak of a \emph{bifurcation delay}. The new threshold is
called the \emph{dynamic oscillation threshold}. Above this threshold, a
periodic regime is established whose amplitude is given approximately by the 2-branch part of the static bifurcation diagram.

The article focuses on providing the necessary elements to calculate
the amplitude envelopes in different conditions, including when the time-variation of a parameter abruptly
changes rate.

\subsection{Similarities and differences between static and dynamic parameter cases}

The duration of the transient is mainly characterized by two aspects:
\begin{itemize}
\item The time constant of the exponential approach or departure from
  the invariant manifold, which is proportional to
  $\log\left(G'(x^*)\right)$, as shown by Eq.~\eqref{sol_stat}
\item the value of the initial condition of $x$, or how far it is from
  the invariant curve or the fixed point.
\end{itemize}
Figure \ref{figEDA:ExTrasCI1ArtRec1} illustrates how, for a similar
exponential time constant (and parameters that are constant in time),
it is possible to obtain very different transient times by changing
the value of the initial conditions.

These are important results for understanding the behaviour of
the system in a situation where the parameters change. The differences
in dynamic parameter contexts are:
\begin{itemize}
\item If the parameter starts increasing at a value below the static
  oscillation threshold, the system will first undergo an approach to
  the invariant curve, and only beyond
  this value will it start the departure phase. In fact the approach can
    be so dramatic that a visible oscillation is only observed
    far beyond the static threshold.
\item The exponential time-constant varies throughout the growth of
  the parameter, but it is not simply given by
  $\log\left(G'(x^*(t))\right)$ at each time $t$.
\end{itemize}

In realistic experimental situations, however, stochastic fluctuations prevent the
system from coming too close to the invariant curve in the approach
phase, and this can reduce the bifurcation delay.




\section{Envelopes for dynamic-parameter regimes}
\label{sec:envel-after-dynam}

This section provides a method to describe the oscillation amplitude
in the particular case of a clarinet model system in which the blowing
pressure parameter increases with time at a small constant
nondimensional rate
$\epsilon \ll 1$:

\begin{subnumcases}{\label{dynsys_ppAR}}
x_n=G\left(x_{n-1},\gamma_n\right)\label{dynsys_ppAR_a}\\
\gamma_{n}=\epsilon n+\gamma_0.
\end{subnumcases}

\subsection{Unlimited precision (noiseless)}
\label{sec:ideal-case-}

First, the case with an arbitrarily high precision is
analysed. $x_n$ is the state variable of the system described in
section~\ref{sec:elementary-modelRe}. With the
knowledge of $x_n$ and its previous value $x_{n-1}$ all remaining
variables of the system can be calculated. In \cite{BergeotNLD2012} it
is shown that during a significant part of a slow transient,  $x_n$ is close to the invariant curve $\phi_{\epsilon}(\gamma)$ described above.

As seen in the previous section, for a constant parameter, the envelope is
well described by an exponential envelope (Eq.~\eqref{sol_stat}),
as long as the state variable $x$ remains sufficiently
close to the fixed point $x^*$ so that function $G$ is well
approximated by its tangent line.


\begin{figure}[h!]
\centering
\includegraphics[width=\columnwidth,keepaspectratio=true]{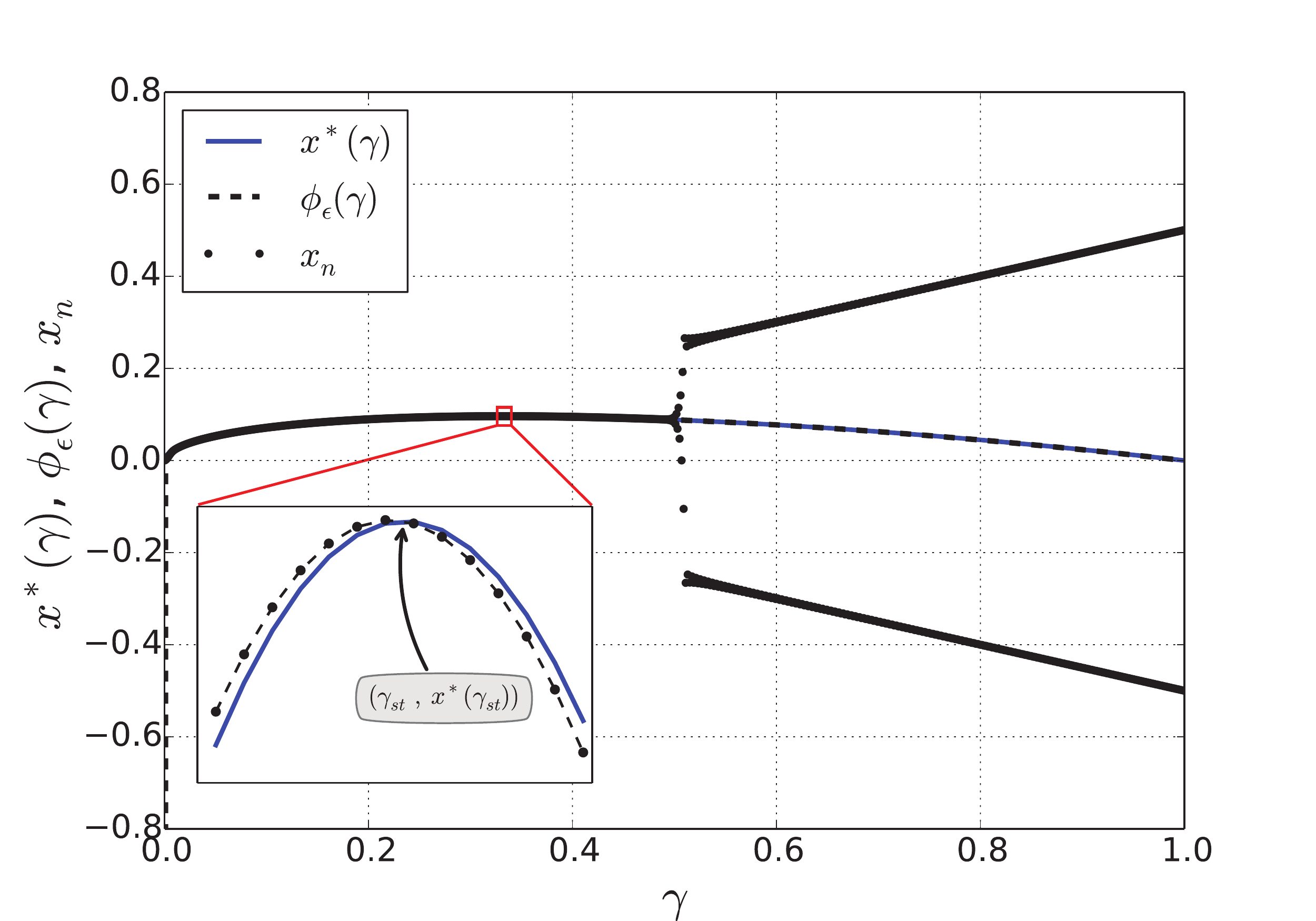}
\caption{(black points) Numerical simulation of the system~(\ref{dynsys_ppAR}). (dashed black line) Invariant curve $\phi_{\epsilon}(\gamma)$. (blue line) Curve of fixed points $x^*(\gamma)$. $\zeta=0.5$, $\epsilon=10^{-3}$ and $\gamma_0=0$.}
 \label{zoom1}
\end{figure}

\rem{BB: paragraphe à revoir; enlever fig.~3} Fig.~\ref{zoom1} suggests that when the parameter $\gamma$ varies
over time, $x_n$ follows more closely the invariant curve than the
curve of fixed points ($x^*(\gamma)$). Instead of following the distance to the
fixed point as in Eq.~\eqref{sol_stat}, a new variable $w_n$ is
therefore defined:
\begin{equation}
w_n = x_n - \phi_\epsilon(\gamma_n).
\label{eq:6}
\end{equation}

Note that, when the parameter is constant, the definition \eqref{eq:6} reverts
to $x-x^*$ of Eq.~\eqref{sol_stat}, as can be verified by substituting
$\epsilon=0$ in the perturbation approximation to $\phi_\epsilon$
(see Appendix~\ref{sec:perturb}, Eq.~\eqref{invcurve_series}).

For small amplitudes $w_n$, the function $G$ in Eq.~\eqref{dynsys_ppAR_a} can be expanded as a first-order Taylor series
around the invariant curve.  The advantage of switching to this
description is that future values of the oscillation amplitude $|w_n|$
can be approximated using a simple function $w(\gamma)$ relating to an initial
iteration $w_0$:

\begin{multline}
|w_n|=w(\gamma_n)\approx \\ |w_0|\exp
\left(\vphantom{\frac{1}{\epsilon}\int_{\gamma_0+\epsilon}^{\gamma_n+\epsilon}\ln\left| G'\phi_\epsilon\left((\gamma'-\epsilon),\gamma'\right)\right|d\gamma'}\right.
\frac{1}{\epsilon}\underbrace{\int_{\gamma_0+\epsilon}^{\gamma_n+\epsilon}\ln\left| G'\left(\phi_\epsilon(\gamma'-\epsilon),\gamma'\right)\right|d\gamma'}_{I(\gamma_n+\epsilon)-I(\gamma_0+\epsilon)}
\left.\vphantom{\frac{1}{\epsilon}\int_{\gamma_0+\epsilon}^{\gamma_n+\epsilon}\ln\left| G'\phi_\epsilon\left((\gamma'-\epsilon),\gamma'\right)\right|d\gamma'}\right).
\label{exact_solution_w3Rec}
\end{multline}

Eq.~\eqref{exact_solution_w3Rec} is the equivalent to Eq.~\eqref{sol_stat} for variable
parameters (see \cite{BergeotNLD2012} for details). Function $I$ is defined by:

\begin{equation}
I(\gamma) = \int_{\gamma_{st}}^{\gamma}\ln\left| G'\left(\phi_\epsilon(\gamma'-\epsilon),\gamma'\right)\right| d\gamma'.
\label{in}
\end{equation}

In the applications shown in this article, $I$ is always used as a
definite integral. As a consequence the integration constant, or one of the bounds of the integral $I$ can be defined arbitrarily.  $\gamma_{st}$ is used in this article as a reference point close to the minimum amplitude, although for $\epsilon\ne 0$ the minimum is attained at a slightly lower pressure.


The discrete equivalent of Eq. \eqref{exact_solution_w3Rec} is:
\begin{eqnarray}
|w_n|&=& |w_0|\exp\left(\sum_{i=1}^{n}\ln\left| \partial_xG\left(\phi(\gamma_i-\epsilon),\gamma_i\right)\right|\right),\nonumber\\
&=& |w_0|\prod_{i=1}^{n} \left|\partial_xG\left(\phi(\gamma_i-\epsilon),\gamma_i\right)\right|.
\label{prodint_w3Rec}
\end{eqnarray}

%

The "product form"~\eqref{prodint_w3Rec} highlights that when the
magnitude of $G'$ is smaller
than $1$ in modulus, which happens before the static threshold
$\gamma_{st}$ is reached, $x_n$ approaches the invariant curve. Beyond
this threshold, $x_n$ moves away from the invariant curve, but
initially at a very slow pace, because the logarithm remains close to
$0$. \rem{BB: pas facile à voir sur l'équation. AA: mieux comme ça?}

\begin{figure}[ht!]
  \centering
  \includegraphics[width=0.9\columnwidth,keepaspectratio=true]{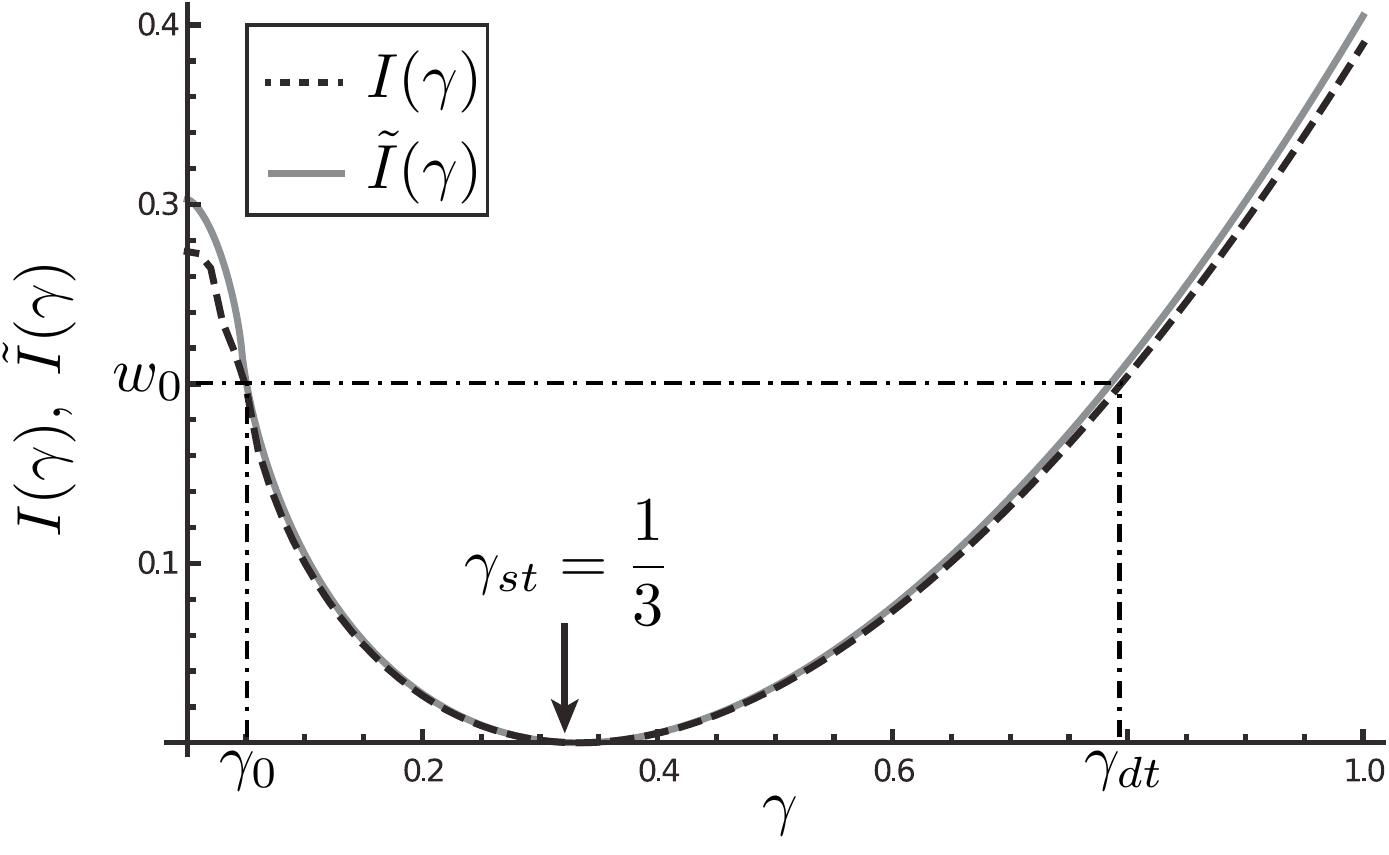}
  \caption{Integral $I(\gamma)$ calculated using Eq.~\eqref{in}
    (dashed line) and approximately using Eq.~\eqref{approx_in} (solid
    line). $\zeta=1/2$, $\epsilon=1/20$.}
  \label{fig:wn05}
\end{figure}

Although $I(\gamma)$ is not easy to calculate analytically, for small values of the increase rate $\epsilon$, the derivative $G'\left(\phi_\epsilon(\gamma'-\epsilon),\gamma'\right)$ can be approximated by its value at the fixed point $G'\left(x^*(\gamma'),\gamma'\right)$, and the integral $I(\gamma)$ written in the form:

\begin{equation}
\tilde{I}(\gamma) = \int_{\gamma_{st}}^{\gamma}\ln\left| G'\left(x^*(\gamma'),\gamma'\right)\right| d\gamma'.
\label{approx_in}
\end{equation}
The error in $I(\gamma)$ committed in this approximation is
observed to be smaller than $\epsilon$ (the difference between
$I(\gamma)$ and $\tilde{I}(\gamma)$ in Fig.~\ref{fig:wn05} is much smaller than $\epsilon$).

For the clarinet model, the expressions involved in the calculation of
the derivative $G'$ are too complicated if function $G$ is used in its
explicit form. However, they can be obtained in a simple form (see
Appendix \ref{sec:perturb}, Eq.~\eqref{eq:17}) from the definition of
$F$ in coordinates $(p,u)$, providing simpler expressions for a
numerical calculation of the integral. In the rest of this paper we
use the approximate form \eqref{approx_in}.

The predicted amplitude $\widetilde{w}(\gamma)$ is calculated as a
distance to the invariant curve:
\begin{equation}
\widetilde{w}(\gamma)=|w_0|\exp\left(\frac{\tilde{I}(\gamma+\epsilon)-\tilde{I}(\gamma_0+\epsilon)}{\epsilon}\right).
\label{eq:Wfunction}
\end{equation}

At iteration $n$, $|w_n| \approx \widetilde{w}(\gamma_n)$. The graphic in Fig.~\ref{fig:wn05} can be used to predict the qualitative behavior of the system: starting at a value $\gamma_0$, the distance to the invariant curve is a monotonic function of $\tilde{I}(\gamma+\epsilon)$. Whenever $\tilde{I}(\gamma+\epsilon)<\tilde{I}(\gamma_0+\epsilon)$, the amplitude is smaller than the starting value. Conversely, when $\tilde{I}(\gamma+\epsilon)>\tilde{I}(\gamma_0+\epsilon)$ the amplitude is higher. $\tilde{I}(\gamma+\epsilon)=\tilde{I}(\gamma_0+\epsilon)$ corresponds to the \emph{dynamic oscillation threshold}, as defined in \cite{BergeotNLD2012}.

The curve described by Eq.~\eqref{eq:Wfunction} is often a good approximation of the envelope for most of the range of the growth parameter, except for large values of $w_n$, which typically arise in 2 situations:

\begin{itemize}
\item In the beginning of the transient, where the iterate $x_0$ can
  be far from the invariant curve, depending on the initial
  conditions. Note that the invariant curve usually diverges for small
  values of $\gamma$, so that even for reasonable values of $x_0$, the
  amplitude $w_0$ can be very large. The region where this curve
  diverges depends on $\zeta$, but is usually well below the static
  threshold $\gamma_{st}$ (see Appendix \ref{sec:pert-terms}).
\item At the end of the transient, where $x_n$ finally escapes from
  the invariant curve.
\end{itemize}

In practice these two situations can be avoided by carefully choosing
the time interval of interest. For example, a few initial iterations
may be calculated exactly using the recursive relation
$x_n=G(x_{n-1})$ until they become sufficiently close to the invariant
curve. In the end of the transient the envelope would not be valid for
other reasons, in particular because the linear approximation in
Eq.~\eqref{eq:Wfunction} is not valid (otherwise the envelope would
grow indefinitely). The prediction $\widetilde{w_n}$ is valid until a
few (3 or 4) iterations before the envelope starts stabilising in the
oscillating branch of the bifurcation diagram.

\subsection{Remarks on very low amplitudes}
\label{sec:infl-prec-or}

The curve $\widetilde{w}(\gamma)$ in Eq. \eqref{eq:Wfunction} often
reaches very small values if the value of $\epsilon$ is sufficiently
small. As a quick example of application, consider a simulation started
at a value of $\gamma$ close to $0$. For this case, Fig. \ref{fig:wn05} shows that
the value of the amplitude at $\gamma_{st}$ is
$\widetilde{w}(\gamma_{st})\simeq
|w_0|\exp\left(-\frac{0.3}{\epsilon}\right)$. In this simulation, $0.3$ is the
difference between the minimum of $I$ (at $\gamma\simeq \gamma_{st} =
1/3$) and the starting value of $I$. For
$\epsilon=1/100$, this means that the minimum amplitude will be
$\exp(-30)\simeq 10^{-13}$. Reducing the increase rate by a factor of
ten ($\epsilon=1/1000$) brings the minimum amplitude down to
the suprisingly low value of $\exp(-300)\simeq 5\times10^{-131}$. In general, the minimum amplitude
reached by the system can be roughly calculated with:
\begin{equation}
  \label{eq:4}
  w_{\textrm{min}} = |w_0| \exp\left(\frac{\tilde{I}(\gamma_{st}) - \tilde{I}(\gamma_0)}{\epsilon}\right)
\end{equation}

A few remarks are suggested by
these extremely low values.

\begin{figure}[t!]
\centering
\includegraphics[width=\columnwidth,keepaspectratio=true]{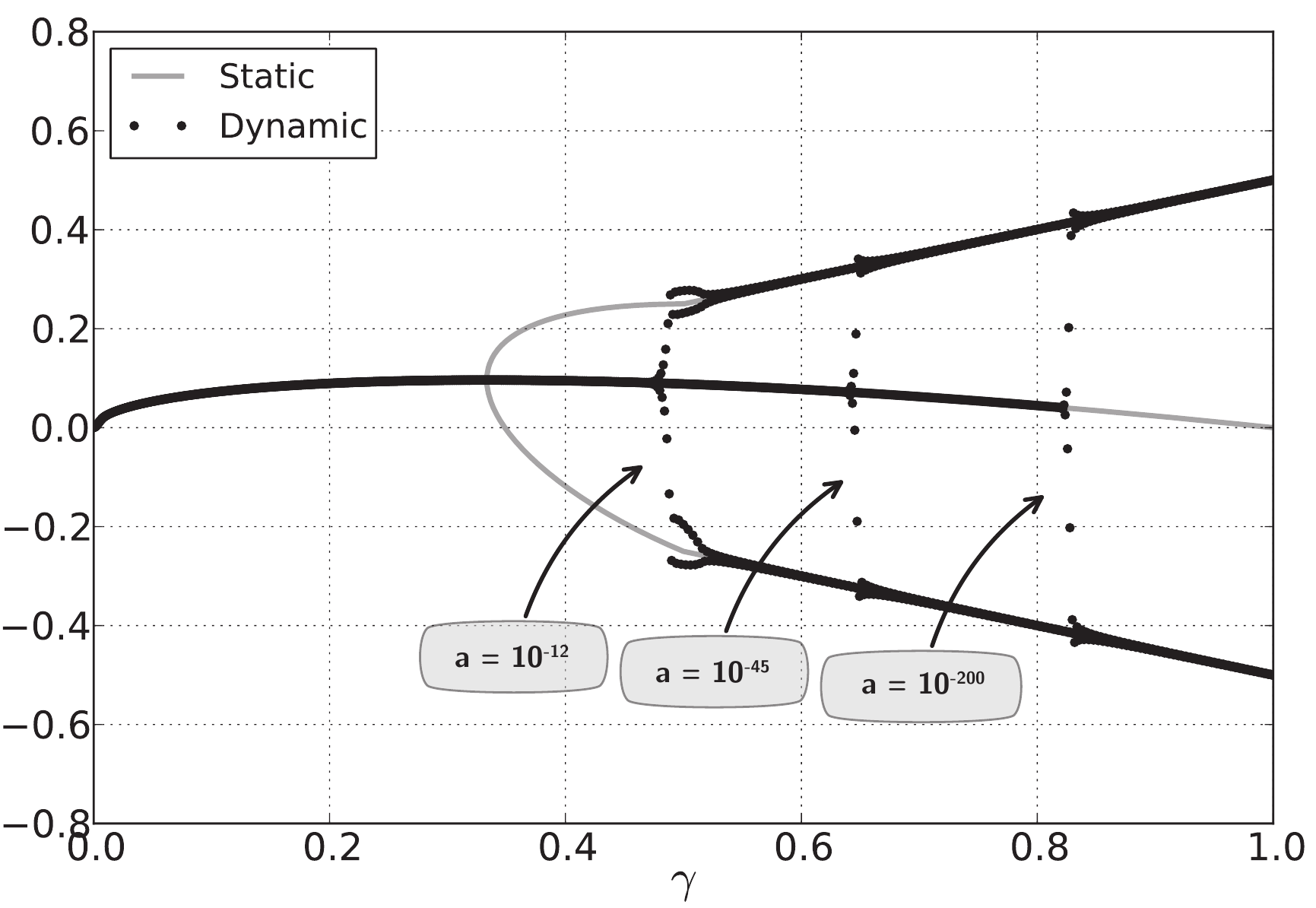}
\caption{Static and dynamic bifurcation for $\zeta=0.5$. Dynamic
  diagram is obtained with $\epsilon=10^{-3}$ and
  $\gamma_0=0$, from numerical
  simulations performed with three different numerical precisions:
  $a=10^{-12}$,$a=10^{-45}$ and $a=10^{-200}$.}
\label{dyn_bif_digramArtRecNo}
\end{figure}

Firstly, extremely low values cannot be
computed using ordinary machine precision. In this article, the
calculations are performed with a \emph{Python} library
(\emph{MPMath}) that simulates arbitrary precision in an ordinary
machine. Fig. \ref{dyn_bif_digramArtRecNo} shows how three different
values of the precision produce very different envelopes. For certain
values of $\gamma$ the errors
are many orders of magnitude higher than
the precision of the calculations. Beyond a certain value of the
precision, the envelope is not greatly affected, only producing
``microscopic'' errors, which are of the same magnitude as the
precision. In practice, the precision $a$ required to simulate the
system should be higher than the
minimum amplitude $w_\textrm{min}$ reached by the system. This ensures
that the difference between the simulation and the exact system never
exceeds $a$, otherwise larger differences are expected because of the
change in dynamic threshold.

Second, even if the simulations are performed using correct
precision, the amplitudes $w$ can only be seen relative to an
accurately calculated invariant curve $\phi$. The estimation of
$\phi_\epsilon$ requires a precision $a_\text{IC}<w_\textrm{min}$ so that it can be used as an accurate reference for determining the amplitudes $w$.

In this paper, the invariant curve is calculated approximately using a
perturbation series (see Appendix \ref{sec:perturb}), whose precision
depends on the number of perturbation terms. Assuming that the
perturbation terms $\phi_i(\gamma)$ all have the same magnitude (which
as shown in Fig. \ref{fig:icterms} is true for $\gamma>1/10$), the
biggest influence in precision comes from the powers of $\epsilon$
that multiply each term in Eq.~\eqref{invcurve_series}. Using this
simple reasoning, a number of terms $n$ is required for an invariant
curve with precision $a_\text{IC}$:


\begin{equation}
\epsilon^n\approx a_\text{IC} \hspace{0.3cm} \Longleftrightarrow \hspace{0.3cm}
n \approx  \frac{\log_{10}(a_\text{IC})}{\log_{10}(\epsilon)}.
\label{eq:CondNumPeTer}
\end{equation}

Returning to the previous example, for $\epsilon=1/100$, $n=7$
perturbation terms are required to observe correctly the envelope $w$
at very low amplitudes, whereas for $\epsilon=1/1000$ the number of
terms is $n=65$. However, even though the invariant curve requires a
lengthy calculation in order to serve as a reference for the
observation of $w$, a direct estimation $\tilde{w}$ can be obtained
with a much cruder approximation of the invariant curve, as shown below.


Note that the previous argument is typically valid for high values of
$\gamma$. For low values, some of the perturbation terms can reach
values higher than 1, especially for high values of $\zeta$. The
argument seems valid in general above the static threshold (see
appendix~\ref{sec:pert-terms} and Fig.~\ref{fig:icterms}).





In real systems, the problem of precision does not
apply. However, experimental systems are very often affected by noise from different
sources. The major source of noise in the clarinet is
turbulence, which cannot be avoided even with a very precise control
of the pressure. Noisy situations, as well as finite precision
situations, can be analysed introducing a stochastic variable in the
iterative system (Eq.~\eqref{dynsys_ppAR})


\subsection{Trajectory of the system affected by noise}
\label{sec:traj-syst-affect}

 If numerical simulations are run with a precision coarser than the $w_\textrm{min}$
 calculated through Eq.~\eqref{eq:4}, the previous formul{\ae}
 must be extended. The limited precision
 (i.e. below $w_\textrm{min}$) used in simulations is modelled as a stochastic variable (with
 a standard deviation of $\sigma=a$) in the system. This case is
 studied in~\cite{BergeotNLD2012b}. A ``squared average'' trajectory
 $<w_n^2>$ is described by:




\begin{multline}
  <w_n^2> \ \approx \underbrace{\widetilde{w}(\gamma_n)^2}_{A(\gamma_n)}+
  \underbrace{\frac{\sigma^2}{\epsilon}
    \int_{\gamma_0+\epsilon}^{\gamma_n+\epsilon} \left(\frac{\widetilde{w}(\gamma_n)}{\widetilde{w}(\gamma')}\right)^2d
    \gamma'}_{B(\gamma_n)}.
  \label{eq:2}
\end{multline}


The two terms of the right-hand side of Eq.~\eqref{eq:2} are functions
of the parameter $\gamma$. The term labeled $A(\gamma)$ corresponds to
the approximation of the trajectory in the absence of noise, the same
as in Eq.~ \eqref{exact_solution_w3Rec}. $B(\gamma)$ is the expected
value of the additional distance to the invariant curve due to the
presence of noise. In practice, when the noise level is sufficiently
high or the precision low (relative to the estimation of
Eq.~\eqref{eq:CondNumPeTer}), only the term $B(\gamma)$ is relevant,
i.e. the trajectory of the system is described by:
$
  \sqrt{<w_n^2>}  \ \approx \sqrt{B(\gamma_n)}
$
with

%
%

\begin{equation}
  \label{eq:bn}
  B(\gamma) = \frac{\sigma^2}{\epsilon}\int_{\gamma_0+\epsilon}^{\gamma+\epsilon}\left(\frac{w(\gamma)}{w(\gamma')}\right)^2 d\gamma'.
\end{equation}

For $\epsilon$ sufficiently small, since $\tilde{I}>0$ by definition
and considering Eq.~\eqref{eq:Wfunction}, it can be deduced that
$\widetilde{w}(\gamma)\gg \widetilde{w}(\gamma')$  for $\gamma'$ close to $\gamma_{st}$
(keeping in mind that $\widetilde{w}$ depends exponentially on $\tilde{I}/\epsilon$ with
$\epsilon$ small, in this article, and the minimum of $I$ and $w$ are
close to $\gamma_{st}$) and $w$ is negligible for all remaining values of $\gamma'$. This allows a simplification of the expression for
$B(\gamma)$ as described below.

According to the shape of $\tilde{I}(\gamma)$ (see
Fig.~\ref{fig:wn05}), a second-order Taylor expansion of
$\tilde{I}(\gamma)$ around the static oscillation threshold $\gamma_{st}$ is used to
simplify its expression (for details, see Appendix \ref{secan:Itilde}):

\begin{equation}
  \label{eq:wngst}
  \tilde{I}(\gamma) \approx 3\sqrt{3}\frac{\zeta}{2} (\gamma-\gamma_{st})^2.
\end{equation}

Using approximation~\eqref{eq:wngst}, the expression of $B(\gamma)$ can be simplified to:

\begin{equation}
  \label{eq:bngstwn1}
  B(\gamma) = \sigma^2\sqrt{\frac{\pi}{3\sqrt{3}\zeta\epsilon}}\exp\left(2\frac{\tilde{I}(\gamma+\epsilon)}{\epsilon}\right).
\end{equation}

Details of the calculations of the simplified expression
\eqref{eq:bngstwn1} are given in Appendix~\ref{sec:detailBn}. This
amplitude $B(\gamma)$ does not depend on the starting amplitude $w_0$,
and is also independent of the starting value of $\gamma$.

It is interesting to notice that according to \eqref{eq:Wfunction},
expression~\eqref{eq:bngstwn1} can also be written:

\begin{multline}
\label{eq:bngstwn2}
B(\gamma)= \sigma^2\sqrt{\frac{\pi}{3\sqrt{3}\zeta\epsilon}}\\
\times \exp\left(2\frac{\tilde{I}(\gamma_0+\epsilon)}{\epsilon}\right)
\left(\frac{w(\gamma)}{w_0}\right)^2.
\end{multline}

In this form, Eq.~\eqref{eq:bngstwn2} shows that, in the presence of noise and far beyond the static threshold, the envelope followed by the system has the same shape as without noise, but with a different amplitude, i.e in this case we have:

\begin{equation}
\label{eq:bngstwn3}
  \sqrt{<w_n^2>}  \ \approx \sqrt{B(\gamma_n)} \approx K \ w(\gamma_n),
\end{equation}
where $K$ is a constant deduced from Eq.~\eqref{eq:bngstwn2}.

As a remark, a different calculation with similar objectives is made
in a previous article~\cite{BergeotNLD2012b} to determine the dynamic
thresholds in presence of noise. The approximation
\eqref{eq:wngst} was used formally to integrate
$\tilde{I}(\gamma_n+\epsilon)$ in Eq.~\eqref{eq:bngstwn1}. The result
is an explicit expression for $B(\gamma)$, and therefore of the dynamic
oscillation threshold. Here, $\tilde{I}(\gamma_n+\epsilon)$ is
numerically integrated, keeping its precise expression given
by Eq.~\eqref{approx_in}. This leads to a better estimation of the
envelope, but that envelope does not have an analytic expression.




\section{Interrupted variation of the mouth pressure parameter}
\label{sec:interr-ramps-contr}


This section describes the behaviour of the system for an example
profile consisting of a limited linear growth of the parameter at a constant rate $\epsilon$ followed by a constant value $\gamma_M$ for an indefinite period of time. The parameter is therefore formally defined as:

\begin{subnumcases}{\label{Gamma_prof}\gamma_n=}
\epsilon n+\gamma_0 \  \text{if} \ n \le M\\
\gamma_M\  \text{if} \ n > M.
\end{subnumcases}


\begin{figure}[h!]
\centering
\includegraphics[width=\columnwidth]{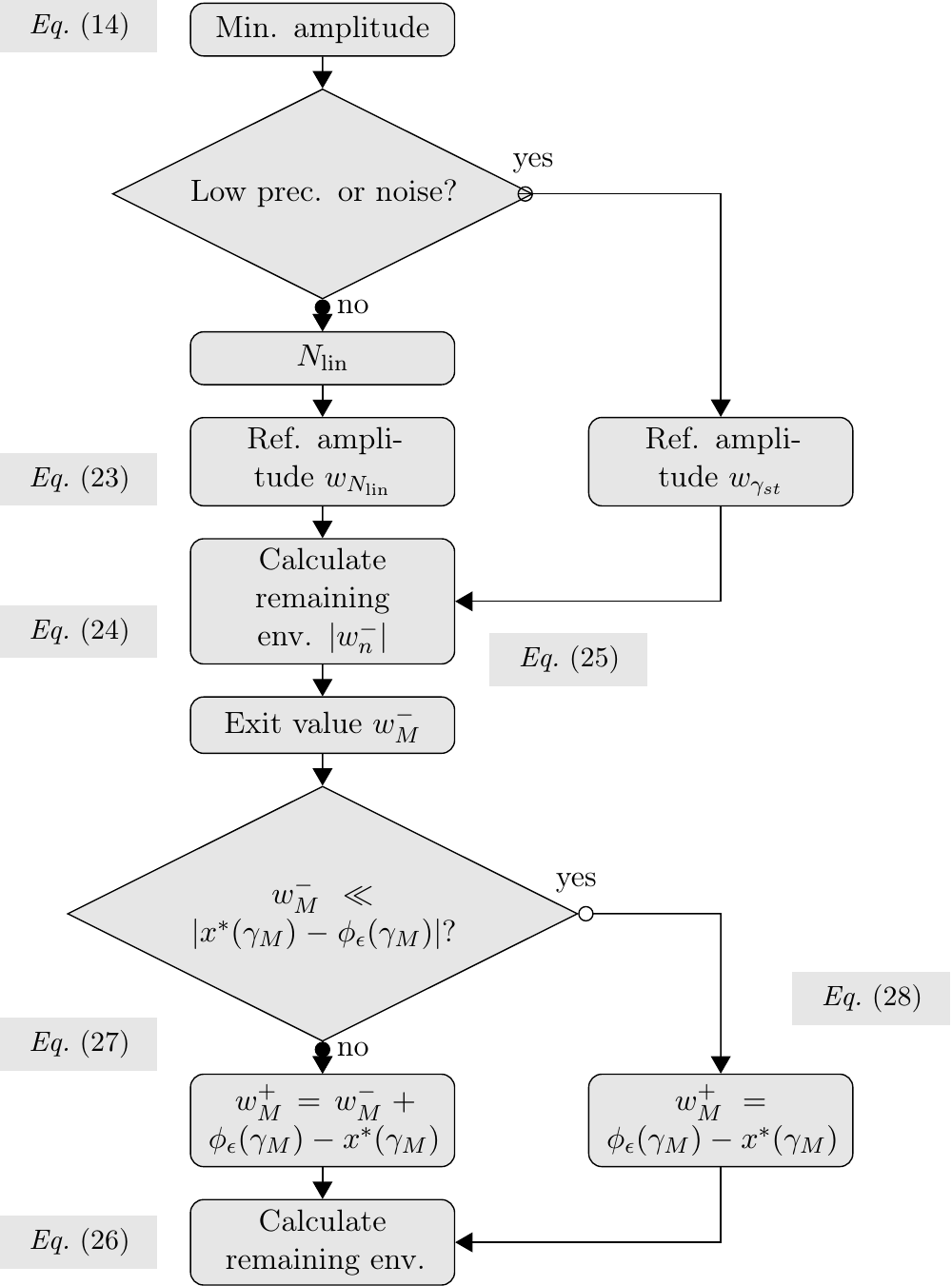}
\caption{Algorithm for determination of the envelope. \rem{Vérifier références, remplacer y/n par yes/no}}
\label{fig:algo}
\end{figure}

Due to the change in increase rate at $n=M$ the growth phase and the
static phase are studied independently. An amplitude envelope
$\tilde{w} ^-(\gamma)$ is computed for the growth phase and another $\tilde{w} ^+(\gamma)$ for the static phase. The two envelopes are connected at $n=M$ since the initial value $\tilde{w} ^+(\gamma_M)$ is deduced from $\tilde{w} ^-(\gamma_M)$. The method is described in the next sections and summarised in Fig.~\ref{fig:algo}.

\subsection{Amplitude envelope of the growing phase: $w^-$}

As explained in section \ref{sec:envel-after-dynam}, the first few
($N_{\text{lin}}$) iterations must usually be performed
manually. These correspond to an ``approach phase'' that brings the
system close enough to the invariant curve so that the assumption of
linearity is valid.

At iteration $N_{\text{lin}}$ the
state of the system is given by:
\begin{eqnarray}
  \label{eq:10}
  n &=& N_{\text{lin}} \nonumber \\
  x_{N_{\text{lin}}} &=& G^{N_{\text{lin}}}\left(x_0\right) = \underbrace{G \circ G \circ \ldots \circ G}_{N_{\text{lin}} \text{times}}(x_0) \nonumber \\
  \gamma_{N_{\text{lin}}} &=& \gamma_0 + N_{\text{lin}} \epsilon\nonumber\\
   w^+_{N_{\text{lin}}}&=&x_{N_{\text{lin}}}-\phi_{\epsilon}(\gamma_{N_{\text{lin}}})
\end{eqnarray}

The number of iterations required for the approach phase depends on the starting value of $\gamma$ and the increase rate $\epsilon$. In practice the state of the system is simulated iteratively until it reaches an amplitude $w_n<\epsilon$.

A complication to this view arises when $\gamma$ goes through a
superstable point $\gamma_{ss}$ defined by
$G'\left(x^*(\gamma_{ss}),\gamma_{ss}\right)=0$. At this point the
iterations can approach arbitrarily the invariant curve. Although
this situation can be analysed under some simplifying assumptions,
this is not done in this article, and the reader is referred to
Baesens \cite{Baesens1991} for a detailed description of this case or
to Bergeot \cite{ThesisBBergeot} in the context of the clarinet. A
simple way of circumventing this problem is to force $N_\textrm{lin}$
to bring $\gamma$ beyond the super-stable point.



A few explicit iterations (usually less than 5) allow the calculation
of the amplitude $w(\gamma_{N_{\text{lin}}})$. Iteration $n=N_\text{lin}$
is used as a safe starting point for the analytic determination of the
envelope.

\begin{multline}
\widetilde{w}^-(\gamma)=\left|w^-_{N_{\text{lin}}}\right|\\
\times \exp\left(\frac{\tilde{I}(\gamma+\epsilon)-\tilde{I}(\gamma_{N_{\text{lin}}} +\epsilon)}{\epsilon}\right).
\label{eq:WfunctionPl}
\end{multline}

When the simulations are performed with a lower precision than that
required for simulating the exact system (see
Eq.~\eqref{eq:CondNumPeTer}), the initial value $\gamma_0$ does not
affect the growth of oscillations. In this case, an average squared amplitude is given by Eq.~\eqref{eq:bngstwn1} starting from $\gamma_{st}$. Therefore, the envelope is given by:

\begin{equation}
  \label{eq:rec-noise}
  \widetilde{w}^-(\gamma) = \sigma \left(\frac{\pi}{3\sqrt{3}\zeta\epsilon}\right)^{1/4}\exp\left(\frac{\tilde{I}(\gamma+\epsilon)}{\epsilon}\right).
\end{equation}

This approximation is valid for $\gamma>\gamma_{st}$, which is the
usual region of interest. Below $\gamma_{st}$ the oscillations are
mostly random, with an average level that remains close to the
standard deviation of the stochastic perturbation $\sigma$.

\subsection{Amplitude envelope of the static phase: $w^+$}
\label{secAR:wplus}


 At $n=M$, $\gamma$ becomes constant and
the oscillation undergoes an exponential growth (provided that
$\gamma_M> \gamma_{st}$), given by Eq.~\eqref{sol_stat} where the
initial value $w_0$ is replaced by the value $w_M^+$ deduced from the previous study of the growing phase:

\begin{equation}
  \label{eq:1}
  \widetilde{w}_n^+ = \left|w_M^+ \left[G'(x^*(\gamma_M),\gamma_M)\right]^{(n-M)}\right|.
\end{equation}

The starting amplitude $w_M^+$ for the static phase is given by
continuity of $x$:
\begin{equation}
  \label{eq:14}
  w_M^+ = \widetilde{w}^-(\gamma_M) + \phi_\epsilon(\gamma_M) - x^*(\gamma_M).
\end{equation}
due to the change in invariant manifold from the invariant curve
$\phi_{\epsilon}(\gamma)$ to $x^*(\gamma_M)$.

As a remark, when the amplitude $ \widetilde{w}^-(\gamma_M)$ is sufficiently small (i.~e. $ \widetilde{w}^-(\gamma_M)\ll|x^*(\gamma_M)-\phi_\epsilon(\gamma_M)|$), the starting amplitude can be given simply by the difference between the invariant curve and the curve of fixed points:

\begin{equation}
  \label{eq:141}
  w_M^+ = \phi_\epsilon(\gamma_M) - x^*(\gamma_M).
\end{equation}

In such a situation, the transient time is roughly given by the time
until the slope discontinuity in the blowing pressure profile, plus a delay
corresponding to the time needed for the oscillations to grow from
$w_M^+$ (independently of $w_M^-$) to the final amplitude. Since the
starting amplitude and the exponential coefficient ($G'(x^*)$ in
Eq.~\eqref{sol_stat}) are independent of the slope of the growth
phase, so is the duration of the transient resulting from the interruption in
the growth. This matches observations on real instruments blown
artificially \cite{Jasa2013BBergeot}.

In any case, the oscillation usually starts very close to the fixed
point $x^*(\gamma_M)$. This ensures that the linear approximation is
valid on a large part of the transient (see
Fig.~\ref{figEDA:ExTrasCI1ArtRec1}).

\section{Examples}
\label{sec:examples}

A few examples of simulations are presented in this section, together
with predictions based on the previous sections, and their
limitations. The ``actual envelopes'' corresponding to the absolute
distance between the iterated values and the invariant curve are plotted
together with the estimation of the envelopes (Eq.~\eqref{eq:2}). In examples presented
in sections~\ref{sec:small-gamma_m} and \ref{sec:large-gamma_m} the
numerical precision is higher than the minimum amplitude reached by
the system (Eq.~\eqref{eq:4}). The effect of introducing a stochastic
variable in the system, which plays a similar role as performing
simulations with low precision~\cite{BergeotNLD2012b}, is shown in the
example of section~\ref{sec:EffectNoise}.

\subsection{Interruption below dynamic oscillation threshold}
\label{sec:small-gamma_m}

\begin{figure}[!ht]
  \centering
\includegraphics[width=\columnwidth,keepaspectratio=true]{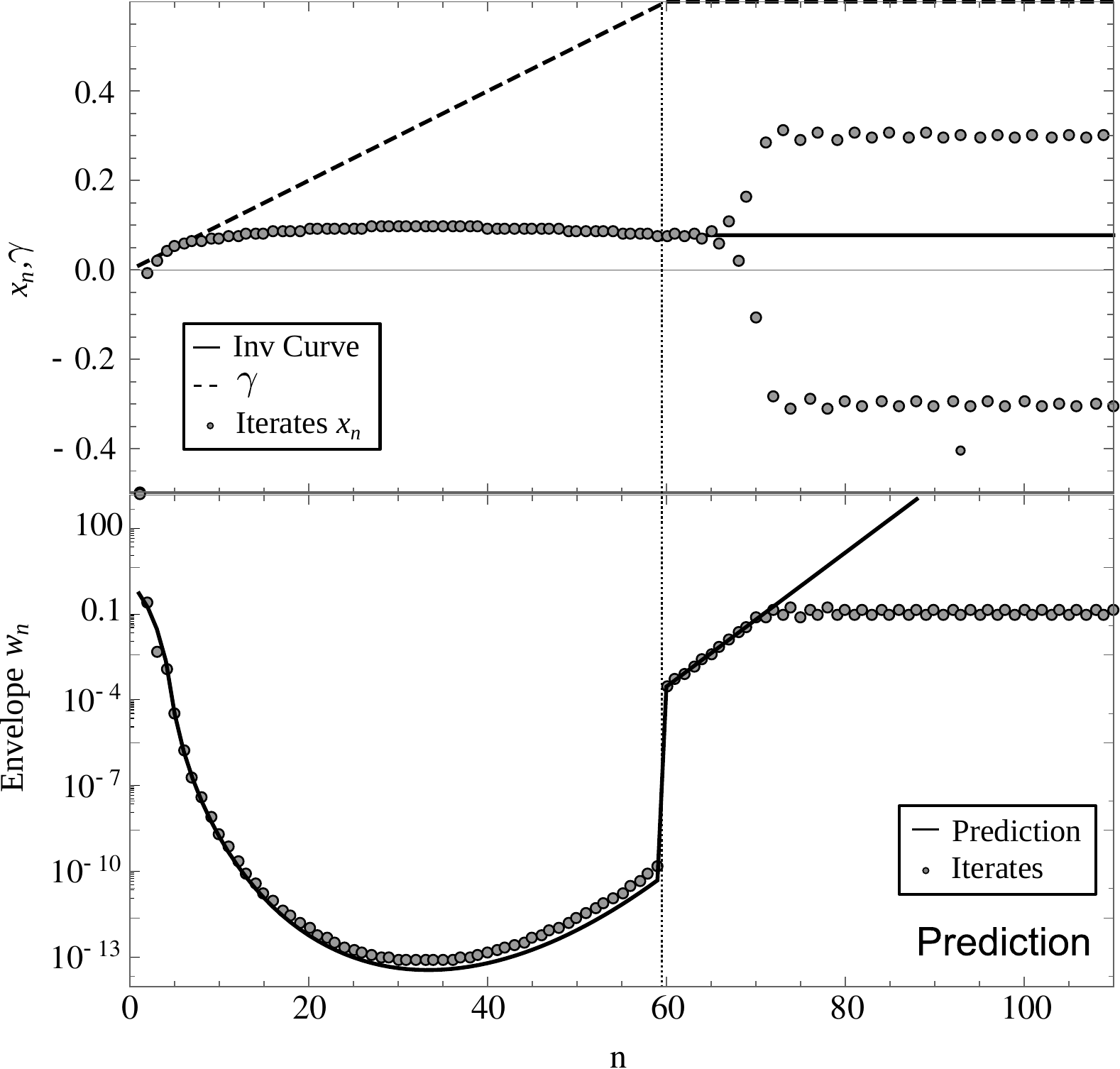}
  \caption{Simulation of the system in Eq.~\eqref{eqdiff_1ArtRec} with
    unlimited precision. The
    invariant curve (Eq.~\eqref{invcurve_1A}) is calculated with 8
    perturbation terms and envelope
    predictions given by Eq.~\eqref{eq:Wfunction}. $\epsilon=0.01$, $\zeta=1/2$, $\gamma_M=0.6$,  $\gamma_0=1/10000$,  $x_0=0.5$.}
  \label{fig:sim06}
\end{figure}

In Fig.~\ref{fig:sim06}, the increase in mouth pressure $\gamma$ is
stopped at a relatively small value of the parameter. In consequence,
the amplitude of the oscillations is considerably smaller when the
increase is interrupted.  A jump in the relative amplitude is observed
when $\gamma=\gamma_M$, in a logarithmic plot (see
Fig.~\ref{fig:sim06}(b)). This jump arises because $w$ is the distance
to the invariant curve $\phi_\epsilon$ before $\gamma_M$ and to the
fixed point $x^*$ after.

In this example, 6 iterations ($N_{\text{lin}}$) are used to reach the
linear approximation. Moreover
$\widetilde{w}^-(\gamma_M)\ll|x^*(\gamma_M)-\phi_\epsilon(\gamma_M)|$,
so that the starting amplitude for the constant parameter phase
($w_M^+$) is deduced from Eq.~\eqref{eq:141}.  The envelope is then computed following the method described above (see Fig.~\ref{fig:algo}).


Fig.~\ref{fig:sim06}(b) also shows that the prediction is slightly in advance relative to the actual envelope. The reason is that $\tilde{I}(\gamma)$ is calculated using a severe approximation $\phi_\epsilon(\gamma-\epsilon)\approx x^*(\gamma)$ (see Eq.~\eqref{approx_in}). For small values of $\epsilon$ the approximation is satisfactory. The advantage of using this approximation is that a single curve $\tilde{I}(\gamma)$ can be used for any small value of the growth rate.

\subsection{Interruption near the dynamic oscillation threshold}
\label{sec:large-gamma_m}

\begin{figure}[!ht]
  \centering
\includegraphics[width=\columnwidth,keepaspectratio=true]{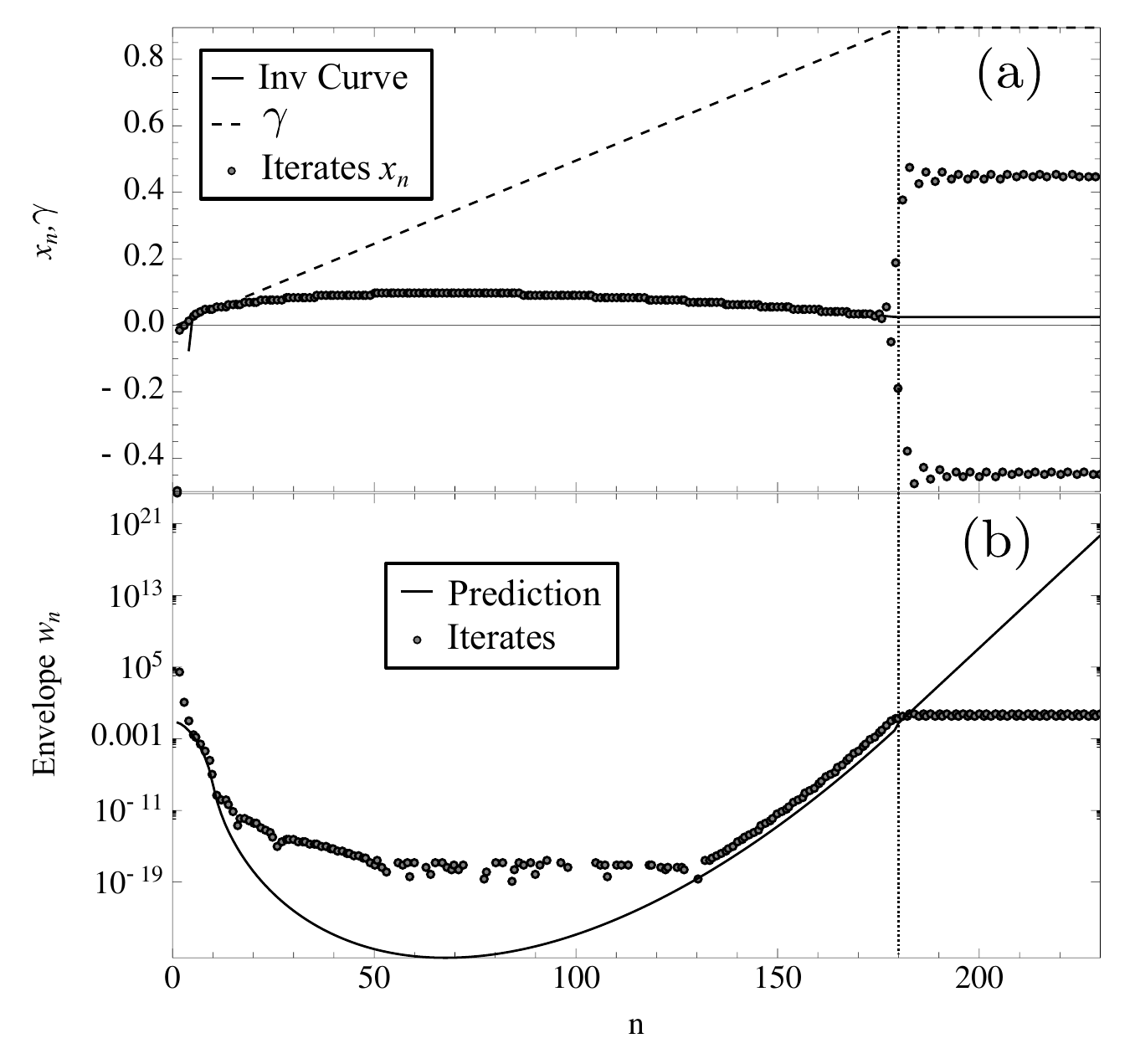}
  \caption{Simulation of the system in Eq.~\eqref{eqdiff_1ArtRec} with
    unlimited precision. The
    invariant curve (Eq.~\eqref{invcurve_1A}) is calculated with 8
    perturbation terms and envelope
    predictions given by Eq.~\eqref{eq:Wfunction}. $\epsilon=.005$, $\zeta=1/2$, $\gamma_M=0.9$, $x_0=0.5$, $\gamma_0=1/10000$.}
  \label{fig:sim09}
\end{figure}

In Fig.~\ref{fig:sim09}, $\gamma$ reaches a higher stable value. This results in higher values of amplitude $w_n$ when the parameter stops increasing.

The envelopes during the growing phase are estimated using the same
method as in the previous example. At iteration $M$, since the system
is estimated to have an amplitude that is higher than the difference
$|\phi(\gamma)-x^*(\gamma_M)|$, the new amplitude $w^+_M$ is the
distance to the invariant curve of the growing phase
$\widetilde{w}^-(\gamma_M)$. The remaining envelope is the exponential  $w^+_M\exp|G'(x^*(\gamma_M),\gamma_M)|$, as in the example above.

The biggest difficulty in estimating the amplitude of the static phase arises when the amplitude $\widetilde{w}^-(\gamma_M)$ is similar to  $|\phi(\gamma)-x^*(\gamma_M)|$. In this case, the iterate at $n=M$ can be either very close to the fixed point curve or at twice the distance between the two reference curves, which will imply very different amplitudes for the static phase.

Finally, in Fig.~\ref{fig:sim09}(b), the jump in relative amplitude at the beginning of the static phase exists but it is not clearly visible because the amplitude of the oscillation at $\gamma=\gamma_n$ is large compared with the case shown in the previous example (see Fig.~\ref{fig:sim06}(b)).


The disagreement between the iterates and the prediction for
$10<n<130$ may appear to suggest that the prediction is not good
here, whereas in fact it is the ``actual envelope'' that is
incorrect. This is due to an inaccurate determination of the invariant
curve. In fact, the number of terms needed for the invariant curve
(Eq. \eqref{eq:CondNumPeTer}) makes its analytical computation too
complicated.
This situation is thus different
from the numerical precision problem outlined in
Fig.~\ref{dyn_bif_digramArtRecNo}, where the iterates are in some
cases very different from those of the ideal system simulated with
infinite precision due to the shift in dynamic threshold. The prediction is valid for most
of the simulation between $n=4$ and $n=160$, and it matches the
envelope whenever the invariant curve is valid (in particular above
$n=130$). This shows that the envelope and the dynamic threshold can
be fairly well predicted, even with an inaccurate approximation.

\subsection{Simulations with noise}
\label{sec:EffectNoise}

\begin{figure}[!ht]
  \centering
\includegraphics[width=\columnwidth,keepaspectratio=true]{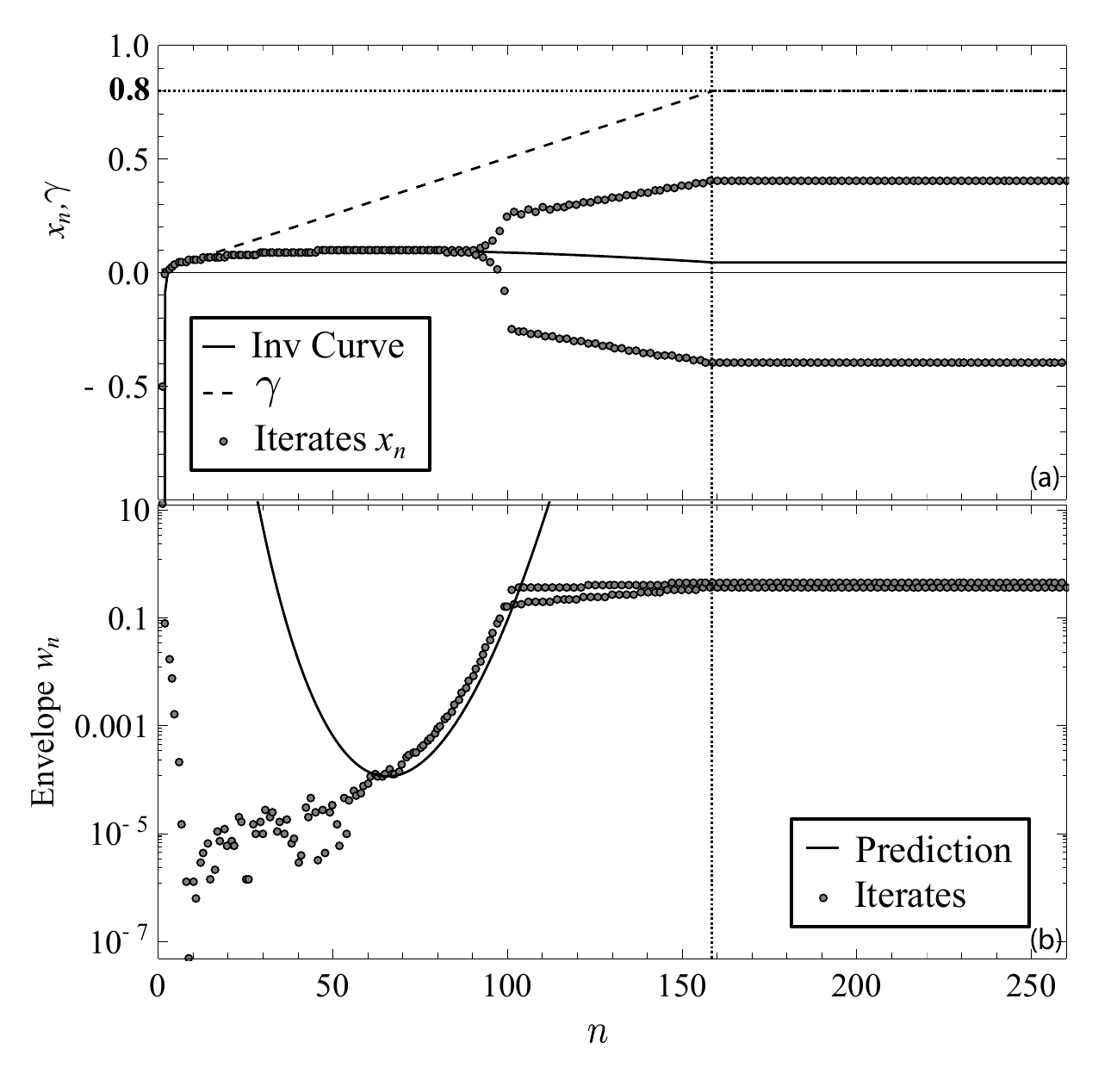}
  \caption{Simulation of the system in Eq.~\eqref{DiffEq_G_InflGamm}
    with linearly increasing $\gamma$ and added noise.
    Invariant curve (Eq.~\eqref{invcurve_1A}) and envelope
    predictions given by Eq.~\eqref{eq:bngstwn3}. Unlimited precision, $\epsilon=.01$, $\zeta=1/2$, $x_0=0.5$, $\gamma_0=1/10$, $\sigma=10^{-4}$.}
  \label{fig:sim-noise}
\end{figure}

In the example of Fig.~\ref{fig:sim-noise}, the simulation is
performed adding a stochastic variable to $\gamma_n$ with a uniform
probability distribution having a standard deviation
$\sigma=10^{-4}$. This is roughly equivalent to a simulation without
noise but with a numerical precision fixed to 4 significant digits
(i.e. $a=10^{-4}$)~\cite{BergeotNLD2012b}.

The sequence $B(\gamma_n)$ (Eq. \eqref{eq:bngstwn1}) is calculated
based on this value of $\sigma$, and its square-root plotted as the
envelope prediction. The prediction is valid for
$\gamma>\gamma_{st}=1/3$, where $B(\gamma_n)$ reaches its minimum
value. The departure of the oscillations occurs earlier when compared
to the case without noise. When $\gamma<\gamma_\text{st}$ the
amplitude is roughly that of the noise, because the noise is added at
each new iteration to a value smaller that of the non-linear function
applied to the previous iteration. \rem{Instead of: Before reaching $\gamma_{st}$ the amplitude
is roughly that of the added random variable, because the random value
is added at each new iteration to a smaller value resulting from the
application of the nonlinear function.}


Because of the reduction in the bifurcation delay, the oscillations
are seen to depart much earlier than the cessation of the increase of the
parameter $\gamma$. After the end of the exponential increase in
amplitude, the envelope increases with the parameter, following the
two-state oscillation given by the static bifurcation diagram.


A discontinuity similar to that observed in figure  \ref{fig:sim06}
can arise also in noisy conditions. However,  because the envelope
curve $w(\gamma)$ multiplies a bigger value at $\gamma_\text{st}$, the
discontinuity is seen only for smaller values of  $\gamma_M$.
\rem{Instead of: When $\gamma$ stabilizes at a low values $\gamma_M$, implying small amplitudes $w^+_M$, the remaining envelope of the recipe given in section~\ref{secAR:wplus} can be applied, giving rise to a similar discontinuity in the envelope as seen in Fig. \ref{fig:sim06}}

Due to the random nature of the system, the prediction should not be interpreted as an approximation to the exact envelope, but rather as the envelope followed on average by a series of runs of the simulation. In fact, in this case, for a series of runs with different noise samples, the actual envelope was seen to shift towards the right or the left by about 4 iterations.

\section{Discussion}


The method can in principle be extended to include frequency
independent losses \cite{dalmont:3294}, although this may be hard to
acheive analitically. The invariant curve cannot be calculated
directly from the simple expression of $F$, as in
Appendix~\ref{sec:perturb}, requiring the use of the much more
complicated expressions of $G$ and its derivatives. More complex
models of clarinets with frequency dependent losses are known to give rise to long attack transients
with similar envelope shapes
\cite{ThFabSil}, and the envelope estimation used in the present article may be similar in models with small
dispersion in the reflection function \cite{Almeida2014}.

When the mouth pressure grows linearly over time, the logarithm of
the amplitude is proportional to a predetermined curve, which
we call $I(\gamma)$. The proportionality factor depends on the inverse
of the growth rate $\epsilon$, whereas the offset depends on one of these two factors:

\begin{itemize}
\item the initial amplitude (starting distance to the invariant
  curve), when the precision is high enough (see
  Eq.~\eqref{eq:Wfunction}) or
\item the stochastic level $\sigma$ when the simulation is imprecise
  or the system is noisy (Eq.~\eqref{eq:bngstwn3}).
\end{itemize}

A stop in the linear growth of the mouth pressure may occur while the
system is still oscillating with low amplitude. In this case, when
the pressure stops increasing, the oscillation resumes exponentially
from a higher amplitude, which is given by the distance between the invariant
curve and the fixed point at the particular value of the mouth
pressure. A discontinuity in the amplitude envelope is observed if
before the mouth pressure stops increasing while the amplitude was still at
a value lower than the distance between the invariant
curve and the fixed point.


Bifurcation delay has also been observed in a real instrument. So
far it has been hard to relate the amplitude envelope to the value of
the mouth pressure. In interrupted ramps of the mouth pressure
however, the oscillations seem to be triggered close to the inflection
point of the blowing pressure~\cite{Jasa2013BBergeot}. In this case,
an exponential amplitude growth then resumes. This is as expected for low
values of $\gamma_M$, as shown in the example in section
\ref{sec:small-gamma_m}.

%

The values of $\gamma$ at the start of the oscillation depend on the rate of
growth of the mouth pressure, an indication that the system is
determined by the stochastic fluctuations in the mouth
pressure.




For a constantly increasing parameter, the \emph{dynamic oscillation
threshold} $\gamma_{dt}$~\cite{BergeotNLD2012,BergeotNLD2012b} gives
the approximate value of the mouth pressure parameter for which an
audible sound appears, or in other terms, the distance from the invariant $w$ curve
becomes ``macroscopic''. When the linear growth of the mouth pressure
is suddenly stopped at $n=M$ and then kept constant at a value
$\gamma_M$, two situations must be distinguished:

\begin{enumerate}[$\bullet$]
\item $\gamma_M < \gamma_{dt}$: a growing exponential envelope starts
  at $\gamma=\gamma_{M}$ with a fixed starting amplitude, which
  only depends on the value of $\gamma_M$ (see
  section~\ref{sec:small-gamma_m}). Audible sound occurs at a fixed
  time interval from the stop in pressure increase;
\item $\gamma_M > \gamma_{dt}$: the audible (``macroscopic'') sound
  begins at $\gamma=\gamma_{dt}$ (see sections~\ref{sec:large-gamma_m}
  and \ref{sec:EffectNoise}).
\end{enumerate}

In most practical cases the latter situation is more common:
because of the limited precision or noise, $\gamma_{dt}$ is
effectively reduced to values that are much closer to the static
threshold.



\section{Conclusion}
\label{sec:concl}

This work shows that the amplitude envelope produced with a regular
increase of blowing pressure in a simplified clarinet
system can be described reasonably well by the use of a single
function $I(\gamma)$ that is a characteristic of the system. This
function can be used in exact and ``noisy'' cases to describe the
envelope beyond the static threshold $\gamma_{st}$.

When the pressure increase is interrupted, the exponential envelope
corresponding to the transient of a static-parameter case can be matched
with the one corresponding to growing pressures. In many practical
cases, when the interruption occurs at sufficiently low values of the
blowing pressure, this corresponds to a fixed starting amplitude so
that the transient time measured from the interruption is roughly
independent of the previous history of the system.

These conclusions show some dramatic effects of the stabilisation of
the mouth pressure that are due to the discontinuity in
derivative.

In summary, a sudden cessation in the increase inmouth pressure can have a
large impact in the initial transient of the clarinet if it appears at a
low enough value of mouth pressure. A preliminary comparison with a
smoother stabilisation profiles \cite{Bergeot2014} suggests that
smoother profiles give rise to slower transients. However, because of the
simple mathematic expressions used for the profiles, they are not easy
to compare to the piecewise linear profiles shown in this
article.


\subsection*{Acknowledgement}

This work is part of the research project SDNS-AIMV ``Syst\`{e}mes
Dynamiques Non-Stationnaires - Application aux Instruments \`{a}
Vent'' (ANR-09-RPDOC-022-01) financed by the \emph{Agence Nationale de
  la Recherche}. The authors thank Prof. Joe Wolfe for useful suggestions
and proof reading.


\appendix

\section{Table of notation}
\label{sec:notation-table}

{\centering
  {\small
    \rowcolors{1}{white}{gray!25}
    \begin{tabular}{|p{1.8cm}|p{5.6cm}|}
      \hline
      $p(t)$ & non-dimensional pressure \\
      $u(t)$ & non-dimensional flow \\
      $\gamma$ & mouth pressure parameter \\
      $\zeta$ & embouchure parameter \\
      $G(x)$ & iterative function \\
      $x_n$ & outgoing wave (also $p^+_n$)\\
      $x^{*}(\gamma)$ & fixed points of the function $G(x)$; same as $\phi_{\epsilon=0}$ \\
      $\phi_\epsilon(\gamma)$ & invariant curve (depends on $\epsilon$)\\
      $I_\epsilon(\gamma)$ & ``Base curve'' used in calculations of $\phi$ and $w$ (depends on $\epsilon$)\\
      $\widetilde{I}(\gamma)$ & Approximation to the ``base curve'' $I$, independent of $\epsilon$ \\
      $w_n$ & difference between a simulated $x_n$ and $\phi$\\
      $\widetilde{w}(\gamma_n)$ or $\widetilde{w}_n$ & prediction of $w_n$  \\
      $\epsilon$  &increase rate of the parameter $\gamma$ \\
      $\sigma$  & level of the white noise  \\
      $a$  & numerical precision used in calculations  \\
      $A(\gamma)$ & deterministic contribution to $w$\\
      $B(\gamma)$ & stochastic contribution to $w$\\
      $\gamma_{st}$ & static oscillation threshold \\
      $\gamma_{dt}$ & dynamic oscillation threshold \\
      $M$ & iteration number at which $\gamma$ stops increasing \\
      $\gamma_M$ & target mouth pressure ($\gamma$) \\
      \hline
    \end{tabular}
  }
}

\section{Perturbation methods for the invariant curve}
\label{sec:perturb}

This appendix presents a perturbation method to calculate the invariant curve, using only expressions of function $u=F(p)$ (see Eq.~\eqref{nonlin_carac_2eq_ad}). This has the advantage of producing much simpler expressions than using function $x=G(-y)$ (Eq.~\eqref{DiffEq_G_InflGamm}). Higher order terms are needed only when determining $w_n$ from a simulation. For all other purposes used in this article, the first order term is usually sufficient.

\subsection{Generic forms of the invariant curve.}
\label{sec:pert-invcurve}

The invariant curve satisfies the following equation:
\begin{equation}
\phi_\epsilon(\gamma)=G\left(\phi_\epsilon(\gamma-\epsilon),\gamma\right).
\label{invcurve_1A}
\end{equation}

The perturbation to order $K$ consists in expressing $\phi_\epsilon$
as a series of terms depending on powers of $\epsilon$ (the
perturbation):

\begin{equation}
\phi(\gamma)=\sum_{i=0}^K \epsilon^i \phi_i(\gamma) + o(\epsilon^{K+1})
\label{invcurve_series}
\end{equation}

Both $\phi_\epsilon$ and $G$ are developed in a power series, $\phi_\epsilon$ around $\gamma$
and $G$ around the first term $\phi_0$

The right-hand side of Eq.~\eqref{invcurve_1A} is then, to 2nd order:
\begin{multline}
G\left(\phi_\epsilon(\gamma-\epsilon),\gamma\right) = G\left(\phi_0(\gamma),\gamma\right) + \\
G'\left(\phi_0(\gamma,\gamma\right) \left(\phi_1(\gamma) - \phi_0'(\gamma)\right) \epsilon + \\
\left(\frac{1}{2}\left(\phi_1(\gamma) - \phi_0'(\gamma)\right)\right)G''\left(\phi_0(\gamma,\gamma\right) +\\ G'\left(\phi_0(\gamma,\gamma\right) \left(\phi_2(\gamma)-\phi_1'(\gamma) - \frac{1}{2}\phi_0''(\gamma)\right) +\\
O(\epsilon^3).
\label{eq:g_phi_series}
\end{multline}

By equating expression \eqref{invcurve_series} on the left-hand side and \eqref{eq:g_phi_series}, it is possible to isolate terms on each power of $\epsilon$, and extract expressions for each of the functions $\phi_i$.The first term is nothing but the definition of the fixed point:
\begin{equation}
  \label{eq:phi-0}
  \phi_0(\gamma) = G\left(\phi_0(\gamma)\right).
\end{equation}

Each of the higher order terms is obtained from lower-order ones:
\begin{eqnarray}
  \label{eq:phi_terms}
  \phi_1 &=& \frac{G'\left(\phi_0\right)\phi_0'}
  {G'\left(\phi_0\right)-1}\label{eqAR:ExpPhi1}\\
  \phi_2 &=& \frac{G'\left(\phi_0\right)\left(2\phi_1\phi_0' -
      \phi_1^2-
      \phi_0^{\prime2}\right)}{2\left(G'\left(\phi_0\right)-1\right)}
  +\\
  &&\frac{G''\left(\phi_0\right)\left(2\phi_1-\phi_0''\right)}
  {2\left(G'\left(\phi_0\right)-1\right)}\label{eqAR:ExpPhi2}\\
  \ldots\nonumber
\end{eqnarray}
All functions and derivatives of the functions $\phi_i$ are taken at $\gamma$. As expected, all the derivatives of $G$ are taken at the fixed point $\phi_0$, and this remains true for higher-order terms too.

\subsection{Derivatives of $G$ at the fixed point.}
\label{sec:pert-deriv}

\begin{figure}[!ht]
  \centering
  \includegraphics[width=1\columnwidth]{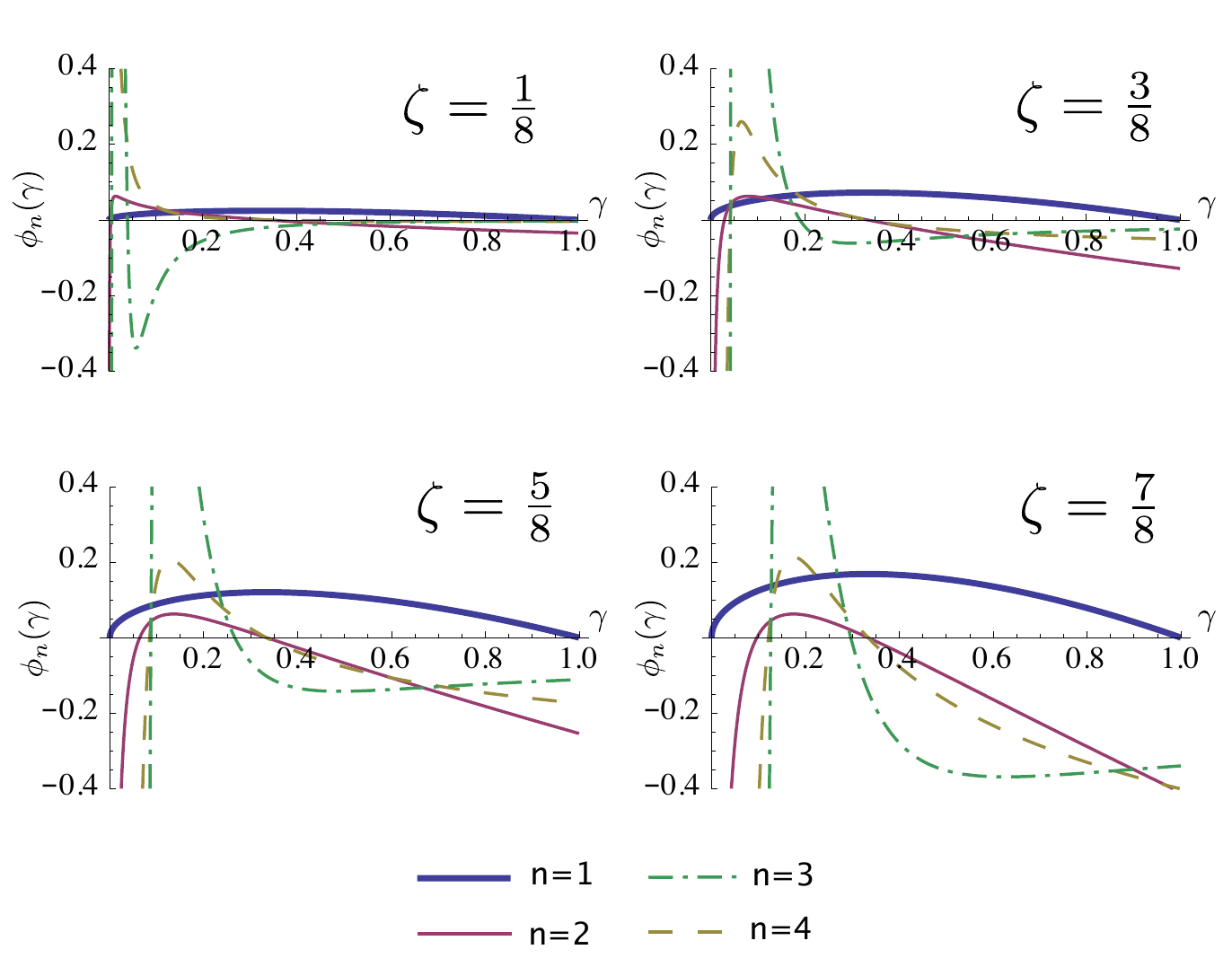}
  \caption{First four terms of the perturbation approximation to the invariant curve as a function of $\gamma$ for four different values of $\zeta$.}
  \label{fig:icterms}
\end{figure}

Derivatives of function $G$, as given by Taillard \cite{NonLin_Tail_2010} are hard to calculate, as the expressions are complex. However, the derivatives of $G$ are related to those of function $F$. $y=G(x)$ can be defined as a parametric curve (with parameter $p$) as the locus of points:
\begin{eqnarray}
  \label{eq:G-par}
  x(p) &=& -\frac{1}{2}\left(p-F(p)\right),\\
  y(p) &=& \frac{1}{2}\left(p+F(p)\right).
\end{eqnarray}
The derivative of the curve $y=G(x)$ is:
\begin{multline}
  \label{eq:17}
  G'=\frac{\frac{\partial y}{\partial p}}{\frac{\partial x}{\partial p}}=
  \frac{F'+1}{F'-1}=\\\frac{-2 \sqrt{\gamma-p}+\zeta\left(3(p- \gamma)+1\right)}{2 \sqrt{\gamma-p}+\zeta\left(3(p- \gamma)+1\right)}.
\end{multline}
All higher order derivatives can be calculated iterativelly:
\begin{equation}
  \label{eq:15}
  G^{(n)}= \frac{\frac{\partial G^{(n-1)}}{\partial p}}{\frac{\partial x}{\partial p}}.
\end{equation}
For instance the second derivative is:
\begin{multline}
  \label{eq:16}
  G''=  -\frac{4F''}{(F'-1)^3} = \\
  -\frac{8 \zeta  (-3 \gamma +3 p-1)}{\left(2 \sqrt{\gamma-p}+\zeta\left(3(p- \gamma)+1\right)\right)^3}.
\end{multline}

In general these formulas are not of much use because they are
functions of $p$ instead of $x$. However, it can be proved that the fixed point of $G$ corresponds to $p=0$ (the line $y=x$ corresponds to the axis $u$), so that:

\begin{equation}
  \label{eqRec:GppF}
  G'(\phi_0(\gamma))=\frac{-2 \sqrt{\gamma}+(1-3 \gamma) \zeta}{2 \sqrt{\gamma}+(1-3\gamma)\zeta},
\end{equation}
and
\begin{equation}
  \label{eq:18}
  G''(\phi_0(\gamma))=-\frac{8 \zeta (3 \gamma +1)}{\left(2 \sqrt{\gamma}+(1-3\gamma)\zeta\right)^3}.
\end{equation}
etc.

\subsection{Perturbation terms.}
\label{sec:pert-terms}

From Eqs~\eqref{eqAR:ExpPhi1} and \eqref{eqAR:ExpPhi2}, the first perturbation terms can be written:

\begin{eqnarray}
  \label{eq:phi0}
  \phi_0(\gamma) &=& \frac{\zeta}{2}(1-\gamma)\sqrt{\gamma},\\
  \label{eq:phi1}
  \phi_1(\gamma) &=& \frac{(1-3 \gamma ) \zeta  \left(\left(3 \gamma -1\right)  \zeta +2 \sqrt{\gamma }
   \right)}{16 \gamma },\\
  \phi_2(\gamma) &=& -\frac{\left(9 \gamma ^2-1\right) \zeta ^2 \left(5\zeta \left(3 \gamma -1 \right)  +8
   \sqrt{\gamma } \right)}{256 \gamma ^{5/2}}.\label{eq:phi2}
\end{eqnarray}

The first perturbation terms ($\phi_i$ for $i=$ 1 to 4) are
represented graphically in Fig.~\ref{fig:icterms}.

\section{Approximated expression of $\tilde{I}(\gamma)$}
\label{secan:Itilde}

According to the shape of $\tilde{I}(\gamma)$, the second-order Taylor expansion of
$\tilde{I}(\gamma)$ around the static oscillation threshold
$\gamma_{st}$ is:

\begin{multline}
  \tilde{I}(\gamma) \approx
  \tilde{I}(\gamma_{st})+(\gamma-\gamma_{st})\tilde{I}'(\gamma_{st})+\\
  \frac{(\gamma-\gamma_{st})^2}{2}\tilde{I}''(\gamma_{st}).
  \label{eqRec:IDL2}
\end{multline}

Through Eq.~\eqref{approx_in}, by definition, we have $\tilde{I}(\gamma_{st})=0$. Since $\tilde{I}(\gamma)$ is the integral of a known function, at the static threshold, the expression of the first and the second derivatives of $\tilde{I}(\gamma)$ are:

\begin{eqnarray}
\tilde{I}'(\gamma_{st}) &=& \ln{\left|G'(x^*(\gamma_{st}),\gamma_{st})\right|} = 0,\\
\tilde{I}''(\gamma_{st}) &=& \left(\frac{d}{d\gamma} \ln{\left|G'(x^*(\gamma),\gamma)\right|}\right)_{\gamma=\gamma_{st}}.
\label{eqRec:Ipp}
\end{eqnarray}

Eq.~\eqref{eqRec:Ipp} can be calculated explicitly from the
expression of $G'(x^*(\gamma),\gamma)$, given by
Eq.~\eqref{eqRec:GppF}. The resulting expression estimated in
$\gamma=\gamma_{st}$. After calculation we obtain
$\tilde{I}''(\gamma_{st}) = 3\sqrt{3}\zeta$, yielding:
\begin{equation}
\tilde{I}(\gamma) \approx 3\sqrt{3}\frac{\zeta}{2} (\gamma-\gamma_{st})^2,
\end{equation}
with a quadratic approximation close to $\gamma_{st}$.

\section{Details of the calculation of the simplified expression of $B(\gamma)$ }
\label{sec:detailBn}

Using Eq.~\eqref{eq:Wfunction}, Eq.~\eqref{eq:bn} is developed,

\begin{eqnarray}
  \label{eq:bngst_0An}
    B(\gamma) &=& \frac{\sigma^2}{\epsilon}\int_{\gamma_0+\epsilon}^{\gamma+\epsilon}\left(\frac{\widetilde{w}(\gamma)}{\widetilde{w}(\gamma')}\right)^2 d\gamma'\nonumber\\
 &=& \nonumber\frac{\sigma^2}{\epsilon} \exp\left(2\frac{\tilde{I}(\gamma+\epsilon)}{\epsilon}\right) \\&&\times \int_{\gamma_0+\epsilon}^{\gamma+\epsilon}
  \exp\left[2\left(\frac{-\tilde{I}(\gamma'+\epsilon)}{\epsilon}\right)\right]d\gamma',
\end{eqnarray}

and replacing $\tilde{I}(\gamma)$ by its expression given by Eq.~\eqref{eq:wngst}, the term $B(\gamma)$ is approximated by:

\begin{multline}
\label{eq:bngstAn}
B(\gamma)  =
\frac{\sigma^2}{\epsilon}\exp\left(2\frac{\tilde{I}(\gamma+\epsilon)}{\epsilon}\right)
\\ \int_{\gamma_0+\epsilon}^{\gamma+\epsilon}\exp\left(-\frac{3\sqrt{3}\zeta}{\epsilon} \left(\gamma'+\epsilon-\gamma_{st}\right)^2\right) d\gamma',
\end{multline}

Eq.~\eqref{eq:bngstAn} can be formally integrated using the error function erf$(x)$~\cite{TaOfIntGrad7thEd}:

\begin{multline}
  \label{eq:bngst2An}
  B_n = \frac{\sigma^2}{\epsilon}\exp\left(2\frac{\tilde{I}(\gamma+\epsilon)}{\epsilon}\right)\frac{1}{2}\sqrt{\frac{\pi \epsilon}{3\sqrt{3}\zeta}}
  \\\times\left[\text{erf}\left(\sqrt{\frac{3\sqrt{3}\zeta}{\epsilon}} \left(\gamma'+\epsilon-\gamma_{st}\right)\right)\right]_{\gamma_0+\epsilon}^{\gamma+\epsilon}.
\end{multline}

The term in square brackets in Eq.~\eqref{eq:bngst2An} can often be
approximated to the values of the error function far from
$\gamma_{st}$, respectively -1 and 1, allowing to write:

\begin{equation}
 \left[\text{erf}\left(\sqrt{\frac{3\sqrt{3}\zeta}{\epsilon}} \left(\gamma'+\epsilon-\gamma_{st}\right)\right)\right]_{\gamma_0+\epsilon}^{\gamma_n+\epsilon} \approx 2.
 \label{eq:SquareBraSimpAn}
\end{equation}





Finally, using Eq.~\eqref{eq:SquareBraSimpAn}, the expression of $B(\gamma)$ becomes:

\begin{equation}
  \label{eq:bngstwn1An}
  B(\gamma) = \sigma^2\sqrt{\frac{\pi}{3\sqrt{3}\zeta\epsilon}}\exp\left(2\frac{\tilde{I}(\gamma_n+\epsilon)}{\epsilon}\right).
\end{equation}


\begin{thebibliography}{10}

\bibitem{McIntyre83:JASA}
M.~E. McIntyre, R.~T. Schumacher, J. Woodhouse: On the oscillations of musical
  instruments.
\newblock J. Acoust. Soc. Am. {\bfseries 74} (Nov. 1983) 1325--1345.

\bibitem{WilJASA1974}
T.~A. Wilson, G.~S. Beavers: Operating modes of the clarinet.
\newblock J. Acoust. Soc. Am. {\bfseries 56} (1974) 653--658.

\bibitem{schum1981}
R.~T. Schumacher: Ab initio calculations of the oscillations of a clarinet.
\newblock Acustica {\bfseries 48} (1981) 71--85.

\bibitem{OllivActAc2005}
S. Ollivier, J.~P. Dalmont, J. Kergomard: Idealized models of reed woodwinds.
  part 2 : On the stability of two-step oscillations.
\newblock Acta Acust. united Ac. {\bfseries 91} (2005) 166--179.

\bibitem{KergoCFA2004}
J. Kergomard, J.~P. Dalmont, J. Gilbert, P. Guillemain: Period doubling on
  cylindrical reed instruments.
\newblock Proceeding of the Joint congress CFA/DAGA 04, 22nd-24th March 2004,
  Strasbourg, France, Soci{\'e}t{\'e} Fran{\c c}aise d'Acoustique - Deutsche
  Gesellschaft f{\"u}r Akustik, 113--114.

\bibitem{NonLin_Tail_2010}
P. Taillard, J. Kergomard, F. Lalo{\"e}: Iterated maps for clarinet-like
  systems.
\newblock Nonlinear Dyn. {\bfseries 62} (2010) 253--271.

\bibitem{KergoActa2000}
J. Kergomard, S. Ollivier, J. Gilbert: Calculation of the spectrum of
  self-sustained oscillators using a variable troncation method.
\newblock Acta Acust. united Ac. {\bfseries 86} (2000) 665--703.

\bibitem{dalmont:3294}
J.~P. Dalmont, J. Gilbert, J. Kergomard, S. Ollivier: An analytical prediction
  of the oscillation and extinction thresholds of a clarinet.
\newblock J. Acoust. Soc. Am. {\bfseries 118} (2005) 3294--3305.

\bibitem{Keefe1992}
D.~H. Keefe: Physical modeling of wind instruments.
\newblock Computer Music Journal {\bfseries 16} (1992) pp. 57--73.

\bibitem{farner2006contribution}
S. Farner, C. Vergez, J. Kergomard, A. Liz{\'e}e: Contribution to harmonic
  balance calculations of self-sustained periodic oscillations with focus on
  single-reed instruments.
\newblock J. Acoust. Soc. Am. {\bfseries 119} (2006) 1794.

\bibitem{Jasa2013BBergeot}
B. Bergeot, A. Almeida, B. Gazengel, C. Vergez, D. Ferrand: Response of an
  artificially blown clarinet to different blowing pressure profiles.
\newblock J. Acoust. Soc. Am. {\bfseries 135} (2014) 479--490.

\bibitem{ThFabSil}
F. Silva: \'{E}mergence des auto-oscillations dans un instrument de musique
  {\`a} anche simple.
\newblock Dissertation.
\newblock Universit{\'e} Aix-Marseille I, 2009.

\bibitem{BergeotNLD2012}
B. Bergeot, C. Vergez, A. Almeida, B. Gazengel: Prediction of the dynamic
  oscillation threshold in a clarinet model with a linearly increasing blowing
  pressure.
\newblock Nonlinear Dyn. {\bfseries 73} (2013) 521--534.

\bibitem{Baesens1991}
C. Baesens: Slow sweep through a period-doubling cascade: Delayed bifurcations
  and renormalisation.
\newblock Physica D {\bfseries 53} (1991) 319--375.

\bibitem{FruchaScaf2003}
A. Fruchard, R. Sch{\"a}fke: Bifurcation delay and difference equations.
\newblock Nonlinearity {\bfseries 16} (2003) 2199--2220.

\bibitem{BergeotNLD2012b}
B. Bergeot, C. Vergez, A. Almeida, B. Gazengel: Prediction of the dynamic
  oscillation threshold in a clarinet model with a linearly increasing blowing
  pressure: Influence of noise.
\newblock Nonlinear Dyn. {\bfseries 74} (2013) 591--605.

\bibitem{OllivActAc2004}
S. Ollivier, J.~P. Dalmont, J. Kergomard: Idealized models of reed woodwinds.
  part 1 : Analogy with bowed string.
\newblock Acta Acust. united Ac. {\bfseries 90} (2004) 1192--1203.

\bibitem{Cha08Belin}
A. Chaigne, J. Kergomard: Instruments {\`a} anche.~-- In: Acoustique des
  instruments de musique.
\newblock Belin, 2008, Ch.~9, 400--468.

\bibitem{ThesisBBergeot}
B. Bergeot: {Naissance des oscillations dans les instruments de type clarinette
  {\`a} param{\`e}tre de contr{\^o}le variable}.
\newblock Dissertation.
\newblock Universit{\'e} du Maine, 2013.

\bibitem{Almeida2014}
A. Almeida, B. Bergeot, C. Vergez: Attack transients in clarinet models with
  different complexity – a comparative view.
\newblock Proceedings of the International Symposium in Music Acoustics, 2014,
  pp. 51--57.

\bibitem{Bergeot2014}
B. Bergeot, A. Almeida, C. Vergez: Effect of the shape of mouth pressure
  variation on dynamic oscillation threshold of a clarinet model.
\newblock Proceedings of the International Symposium in Music Acoustics, 2014,
  pp. 535--540.

\bibitem{TaOfIntGrad7thEd}
I.~S. Gradshteyn, I.~M. Ryzhik: Table of integrals, series, and products (7th
  ed.).
\newblock Academic Press, New York, 1965.

\end{thebibliography}

\end{document}